\documentclass[11pt,a4paper,captions=tableheading,captions=nooneline,DIV=calc]{scrartcl}
\usepackage{amssymb}
\usepackage{amsfonts}
\usepackage{amsmath}
\usepackage{bm}
\usepackage{mathrsfs}
\usepackage{graphicx}
\usepackage{cite}
\usepackage{booktabs}
\usepackage{tabularx}

\setkomafont{caption}{\sffamily\small}
\setkomafont{captionlabel}{\sffamily \bfseries}

\usepackage[textheight=22cm,textwidth=13cm]{geometry}
\begin{document}
\begin{center}

\vspace{2ex}
{\huge\bf Interactive Chemical Reactivity Exploration} \\
\vspace{3ex}

{\large Moritz P.\ Haag$^1$, Alain C.\ Vaucher$^1$, Ma\"{e}l Bosson$^2$, St\'{e}phane Redon$^{2\ast}$, 
Markus Reiher$^{1\ast}$} \\ 
\vspace{1ex}
$^1$ ETH Z\"urich, Laboratorium f{\"u}r Physikalische Chemie, \\ 
Vladimir-Prelog-Weg 2, CH-8093 Z\"urich, Switzerland \\
$^2$NANO-D - INRIA Grenoble - Rh\^one-Alpes/CNRS Laboratoire Jean Kuntzmann, 655, avenue de l'Europe 
Montbonnot, 38334 Saint Ismier Cedex, France \\[2ex]
\end{center}

\vfill
\begin{center}\it Abstract \end{center}

Elucidating chemical reactivity in complex molecular assemblies of a few hundred atoms is, despite 
the remarkable progress in quantum chemistry, still a major challenge. Black-box search methods to 
find intermediates and transition-state structures might fail in such situations because of the 
high-dimensionality of the potential energy surface. Here, we propose the concept of interactive 
chemical reactivity exploration to effectively introduce the chemist's intuition into the search 
process. We employ a haptic pointer device with force-feedback to allow the operator the direct
manipulation of structures in three dimensions along with simultaneous perception of the quantum 
mechanical response upon structure modification as forces. We elaborate on the details of how such 
an interactive exploration should proceed and which technical difficulties need to be overcome. All 
reactivity-exploration concepts developed for this purpose have been implemented in the {\sc Samson} 
programming environment. 

\vfill

\begin{center}
\textit{\today}
\end{center}

\vfill

$\ast$ Corresponding authors:\\
S.~Redon (stephane.redon@inria.fr) and \\
M.~Reiher (markus.reiher@phys.chem.ethz.ch)

\newpage



\section{Introduction}

Unraveling reaction mechanisms on an atomistic level is one of the major goals in quantum chemistry. 
Reactivity studies require calculations based on the first principles of quantum mechanics to 
describe the breaking and forming of chemical bonds in a general way not restricted to certain 
classes of molecules. The algorithmic developments in density functional theory (DFT) and in 
Hartree--Fock-based electronic structure methods have enabled chemists to calculate the electronic 
structure of molecular systems up to several hundred atoms with sufficient accuracy and in a 
reasonable time\cite{dykstra2005}.

A huge effort was made in the past years to allow for the calculation of larger and larger 
molecules\cite{goedecker1999,rubensson2011,ochsenfeld2007}. Despite this success, it is still a 
major task to explore reaction mechanisms for molecular systems of even medium size (say, one to a 
few hundred atoms). This is mainly because of the fact that the unsupervised automated exploration 
of the first-principles potential energy surface (PES) of reactive molecular assemblies is most 
often prohibitive. The trial-and-error approach currently applied (guessing the important structures 
and refining them with local optimization methods) requires experience, luck, and time. The need to 
compose lengthy and at times cryptic text-based input files and to set up large three-dimensional 
structures with input devices working in only two dimensions like the computer mouse, further hamper 
a simple and intuitive application.

A more convenient set-up of molecular structures and default values for electronic structure 
programs is desirable (and in parts available by standard graphical user interfaces). Elegant 
examples are the structure editors {\sc Samson}\cite{bosson2011} and {\sc Avogadro}
\cite{hanwell2012}. Nevertheless, it remains a cumbersome procedure to explore reaction mechanisms 
in systems of a few hundred atoms with two-dimensional input devices and automated search 
algorithms. 

Haptic Quantum Chemistry\cite{marti2009,haag2011}, Interactive Quantum Chemistry\cite{bosson2012,
bosson2013} and Real-time Quantum Chemistry\cite{haag2013,haag2014a} offer new alternative 
approaches to study reactivity in large three-dimensional molecular systems by providing an 
instantaneous response of the system to the structural manipulation.

A so-called haptic pointer device fulfills two functions as an input device and for transmitting the 
response to the operator. With such a device the operator can perform structure manipulations 
directly in three dimensions in order to probe the reactivity of a molecular assembly. Then, the 
response to this probing can be instantaneously presented to the operator as forces rendered by the 
force-feedback functionality of the device. 

In previous work on haptic interaction with molecular systems\cite{brooks1990,krenek1999,harvey2000,
comai2009,davies2009} the rendered forces where obtained from classical force fields or model 
potentials. As these potentials are not sufficiently general and flexible to account for any kind of 
bond-breaking event, quantum chemical methods based on the first principles of quantum mechanics are
required to calculate the forces.

In this work we present concepts for the intuitive and interactive exploration of chemical 
reactivity with force-feedback devices. We first describe in section \ref{sec:manipulation} how the 
operator-driven manipulation of molecular systems takes place in the study of chemical reactivity 
with haptic pointer devices. This is then followed by an in-depth analysis of problems emerging 
during such manipulations in section \ref{sec:issues}. The implementation of these concepts as 
`Apps' loaded into the {\sc Samson} environment is then described in section \ref{sec:implementation}. We 
also present an App to calculate quantum mechanical energies and forces based on the standard 
non-self-consistent density-functional tight-binding method (DFTB). With this App we are able to 
describe a larger range of molecules than it is possible with the ASED-MO method provided by {\sc Samson}. 
Finally, we summarize our findings and provide an outlook for further extensions and features useful 
for studying chemical reactivity interactively.

\section{Manipulating Molecular Systems with Haptic Pointer Devices}
\label{sec:manipulation}

The term `manipulation' of a molecular system shall denote a manual change of atom positions in a 
molecular assembly. The operator applies structural changes by selecting and moving one or more 
atoms. 

A manipulation creates a sequence of structures that can be interpreted as a path through the
configuration space of the molecular assembly. Providing an energy for each structure then creates a 
reaction energy profile. This one-dimensional profile is a slice through the high-dimensional PES
of the system. The goal of chemical reactivity studies is to find (minimum-energy) paths 
corresponding to the possible reactions of a molecular assembly. They are defined by the 
steepest-descent path from the transition-state structure to the reactants structure or to the 
products structure. Starting from a local minimum structure (minimum on the PES) a structural 
change will result in gradients several (in general, on all) atoms of the system. We may define the 
set of nuclear gradients as $\{\bm{g}_I\}$, where 
\begin{align}
    \bm{g}_I = \bm{\nabla}_I \, E \,\, ,
\end{align}
and $I$ refers to a specific atomic nucleus in the assembly. The set of forces $\{\bm{f}_I\}$
\begin{align}
    \bm{f}_I = - \bm{g}_I \,\, ,
\end{align}
represents the response of the system (here, exerted on each atom $I$) as a result of a structural 
change. This primary response will be followed by the secondary response if the structure is allowed 
to relax by a constrained structure optimization procedure running in the background. Thus, the 
operator is attracted to local minima of the PES which can help finding minimum energy paths. 
However, pushing a system on a minimum energy path against a gradient energetically up hill is 
conceptually not trivial as shall be discussed in this paper.

Haptic pointer devices such as the Phantom Desktop\cite{sensable} device employed in this work are 
very intuitive input as well as output devices. They allow the user to manipulate a structure in 
three dimensions. At the same time, such devices can render the primary force response required
to intuitively steer uphill motions and to recognize evasive motions. The structural response is 
visualized by the rearranging atoms. Haptic pointer devices enable the manipulation of six degrees 
of freedom, i.e., of position and orientation of only one object at a time --- either a single atom 
or several atoms joined as a rigid body. Only the (net) force corresponding to this object can be 
rendered. For the presentation of the forces on other atoms force arrows can be displayed (or the 
operator is forced to switch between atoms or to employ two haptic pointer devices).

\subsection{Coupling the Device to Atoms}

Before we delve into the scientific aspects of interactive reactivity exploration, we should make a
few technical comments on how the device is connected to the molecular events. 

An intuitive way to manipulate a molecular system employing a haptic pointer device is to select an 
atom by positioning the pointer on top of it, clicking a button and then moving the pointer. The 
atom is then expected to follow the position of the haptic pointer. At every new structure which the 
operator is generating by moving some of the atoms a new electronic structure (wave function and
energy) is optimized and new forces are calculated. 

If the atoms are allowed to relax upon the manipulation, the operator's input and the relaxation 
procedure need to be combined. This leads to a competing process between the input and the 
relaxation procedure resulting in the manipulated atoms jumping back and forth. Depending on how 
often the operator input is adopted during the optimization the active atoms follow more or less 
closely the input position. The rapid jumps between the position dictated by the input device and 
the one given by the optimization procedure is, however, problematic, because they can lead to 
sudden force changes and therefore to an instability in the force rendering.

A more stable scheme is to attach the input device only indirectly to the atoms\cite{bolopion2010a}. 
This may be achieved by coupling the input device with a spring to the selected atoms. Thereby, we 
may add an additional force calculated from the distance between the atom position $\bm{r}_I$ and 
the device position $\bm{r}_\mathrm{device}$ to the original force $\bm{f}_I$ of the selected atom 
$I$ during the optimization process
\begin{align}
\bm{f}_\mathrm{device} = \bm{f}_I + k \, ( \bm{r}_I - \bm{r}_\mathrm{device} )  \,\, .
\end{align}
The force constant $k$ can be chosen such that the selected atom follows the input device position 
closely. Still, the manipulated atom position will never be exactly at the position given by the 
input device. In such a coupling scheme it is more convenient to render the force exerted by the 
spring than the force of the connected atom. Then, the force felt by the operator is not directly the 
force that describes the atom's reactivity. 

Such an additional (fictitious) layer thus impairs the intuitive reactivity perception. Therefore, 
it is desirable to have the input device exactly determining the position of the manipulated atoms.
This allows one to have a precise control over the atom positions, which is of paramount importance 
for the study of the subtle features of PES. In this case, the relaxation procedure is subject to a 
constraint imposed by the input device as it optimizes only the inactive (not manipulated) part of 
the molecular assembly. 

\subsection{Categories of Manipulation}
\label{sec:manipcats}

\begin{figure}[h]
    \centering
    \includegraphics[width=0.80\textwidth]{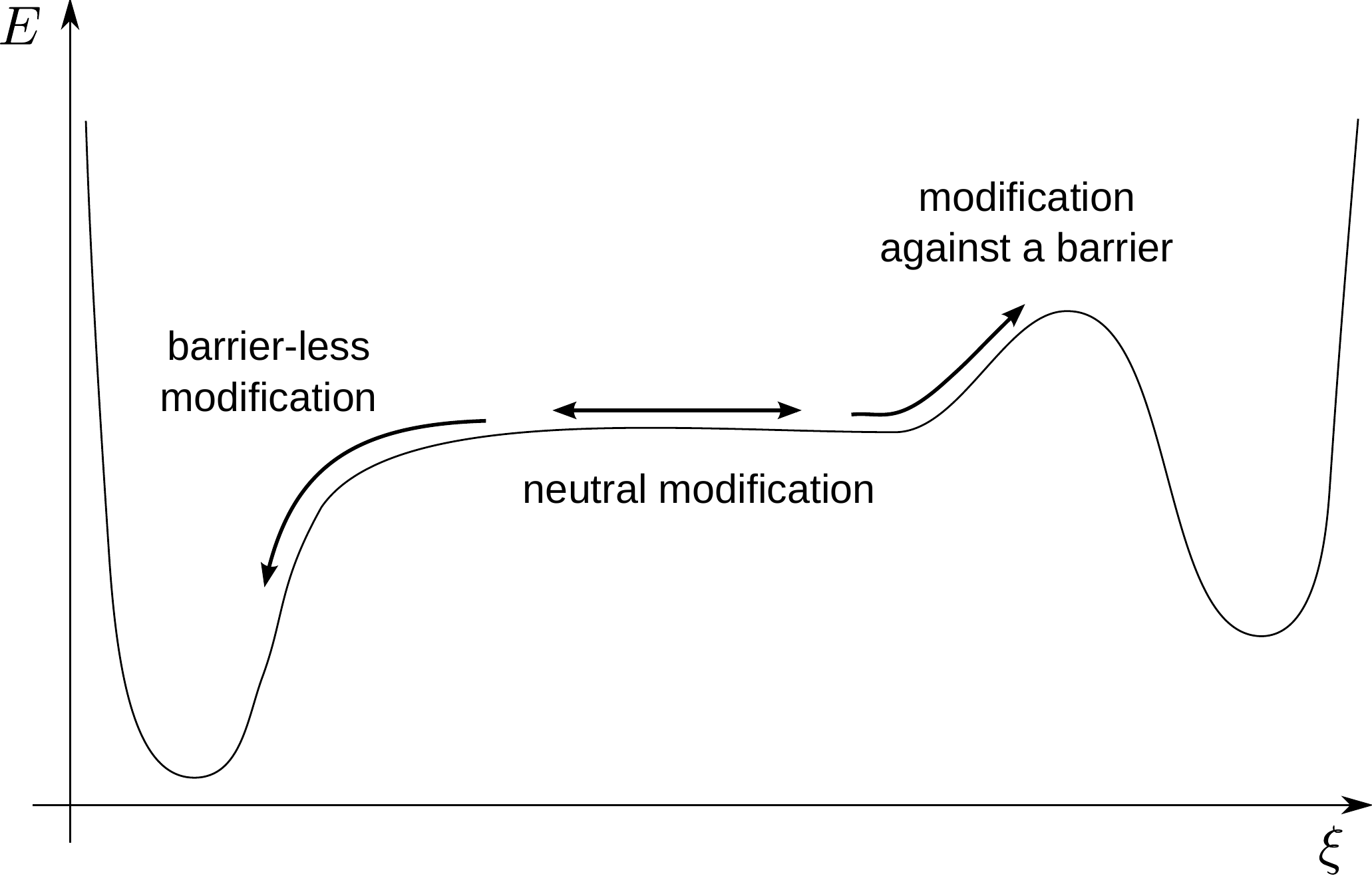}
    \caption{Manipulations categorized according to the energy path profile. The energy $E$ is 
    plotted along a one-dimensional (collective) reaction coordinate $\xi$. Orthogonal coordinates
    leading to an energy increase and thus to restoring forces in these directions are not shown.}
	\label{fig:manip-cat}
\end{figure}

The energy profile of a path described by a structure manipulation in a molecular system may be 
assigned to one of three dominating idealized shapes. Either a path is going uphill in energy with a 
positive gradient (modification against a barrier), downhill with a negative gradient (barrier-less 
modification), or the path is flat with a gradient length of basically zero (neutral modification) 
as depicted in Figure \ref{fig:manip-cat}. The latter case also represents the situation at the top
of a barrier. We chose the one-dimensional reaction coordinate $\xi$ to be always positive in the 
direction of progressing manipulation. Then, a positive gradient corresponds to a negative force 
acting against the direction of manipulation and vice versa. 

We define a \emph{neutral manipulation} as a manipulation by which a property which is part of the 
response shows no detectable change. This implies, however, that for every property that is 
presented to the operator a threshold needs to be defined which defines which numerical values of 
this property are detectable. Note that both the resolution of the rendering and the operator's 
`sensitivity' to the targeted sense determine the detectability. In the case of the haptically 
rendered forces the threshold is determined by the smallest force difference that the human haptic 
sense can detect. Obviously, one can always magnify differences to such an extent that the operator 
will eventually see or feel them if deemed useful. 

Since neutral manipulations do not involve any change in the gradient or energy, they always start 
and also end at stable configurations. Such manipulations will occur frequently in the exploration 
of chemical reactivity, especially at the very beginning of an haptic exploration, if the reactants 
are far apart from each other. Then, the operator can start by orienting and moving the separated 
fragments such that they are in some proper starting position for probing the reactivity without 
experiencing a considerable interaction that would destabilize them. 

Usually a reactive exploration begins by following a contra-gradient path, since the vast majority 
of reaction paths on a PES have a higher lying transition state that needs to be overcome in order 
to reach the product structure.

\section{Two Scales of Reactivity Exploration: Local and Global Exploration}
\label{sec:issues}

Local minima and the interconnecting first-order saddle points describe the reactivity of many  
molecular systems sufficiently well according to Eyring's absolute rate theory\cite{eyring1935} --- 
even if they are as large as enzymes\cite{olsson2006}. To find, describe, and record them is the aim 
of reactivity exploration as developed here. This exploration process can be divided into two 
different parts. The process of finding the structures of reactants and transition states for a 
network of stable intermediates and reaction paths is what we may call \textit{global reactivity 
exploration}. By contrast, exploring a single elementary reaction is in the realm of \textit{local 
reactivity exploration}. From the point of view of global reactivity exploration local reactivity 
exploration is the identification of nodes of the network (local minima) and links between the nodes 
(transition states). Accordingly, both processes are tightly connected and performed simultaneously. 

\subsection{Local Reactivity}

Local minima mark the start and end points of local reactivity exploration. 

\begin{figure}[h!]
    \centering
    \includegraphics[width=0.80\textwidth]{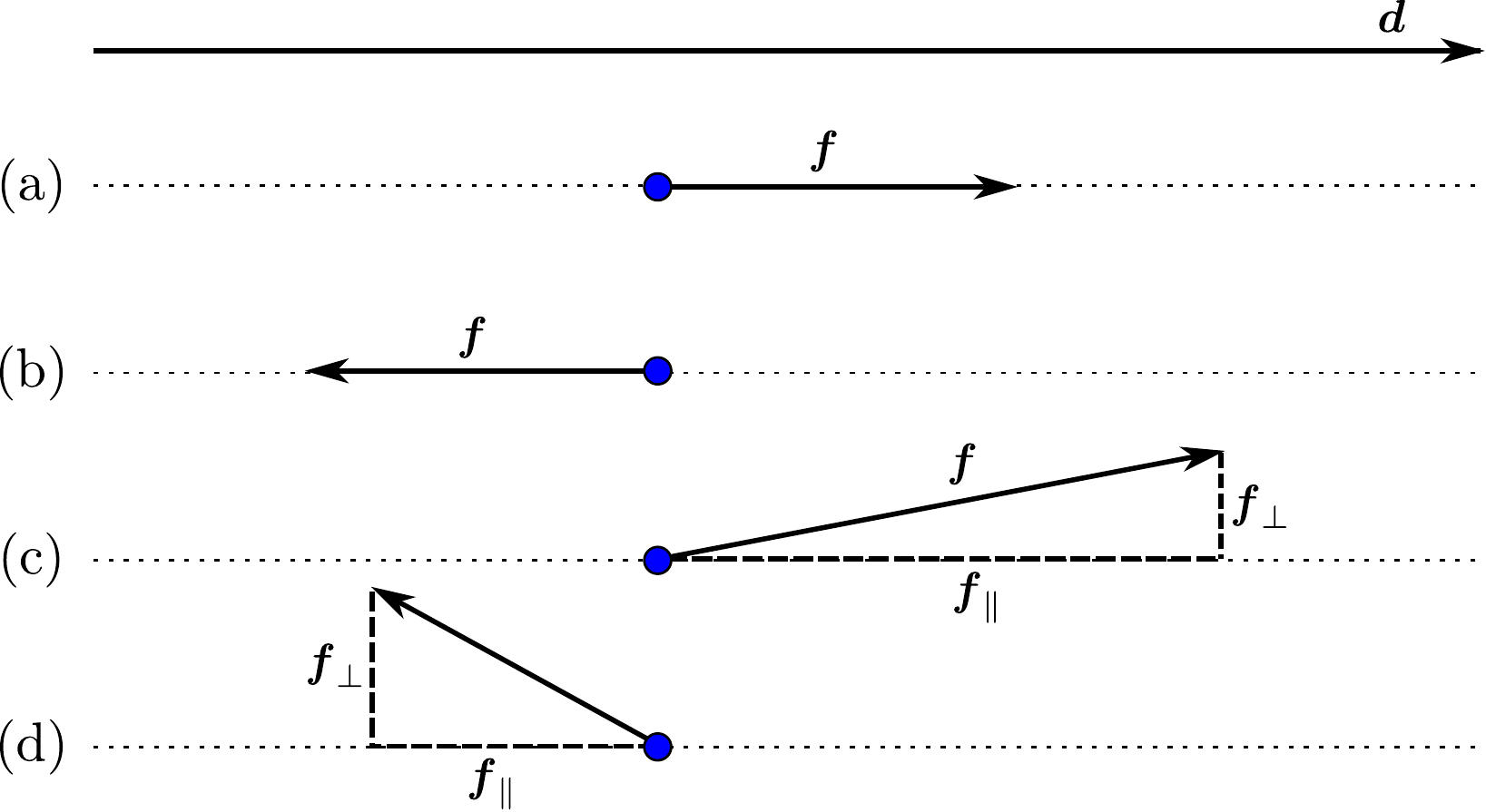}
    \caption{Forces acting on a haptic pointer position during a manipulation. The direction of 
    manipulation $\bm{d}$ and four different cases for the forces $\bm{f}$ are shown. (a) The 
    force points parallel in the direction of the manipulation. (b) The force points parallel but in 
    the opposite direction of the manipulation. (c,d) Two different cases with a perpendicular force 
    component.}
    \label{fig:forces}
\end{figure}

The \textit{intended direction of manipulation} $\bm{d}$ (see Figure \ref{fig:forces}) is a central 
quantity in the following discussion. This direction is estimated by taking the current velocity of 
the manipulation. As soon as the configuration leaves a flat area or a minimum point, gradients 
emerge and the atom controlled by the haptic pointer device experiences a force $\bm{f}$. The 
projection of the force vector onto the intended direction vector is a useful local descriptor with 
which one can classify manipulations. If the projection is positive (see Figure \ref{fig:forces} c), 
the manipulation will be a barrier-less modification. If the projection yields a negative value (see 
Figure \ref{fig:forces} d), the force will be repulsive, i.e., the manipulation is against a 
barrier. 

The projection of the force vector $\bm{f}$ onto the intended manipulation direction vector $\bm{d}$ 
can be further exploited to determine the force components acting parallel, 
\begin{align}\label{eq:forcedecomp1}
 \bm{f}_{\parallel} &= \frac{\bm{f} \cdot \bm{d}}{\left| \bm{d}\right|^2}\, \bm{d} \,\, ,
 \intertext{and perpendicular, }\label{eq:forcedecomp2}
 \bm{f}_{\perp} &= \bm{f} - \bm{f}_{\parallel} \,\, ,
\end{align}
with respect to the intended direction. The components can now be monitored during an exploration 
and they may be scaled independently if deemed useful. 

Decisive for the operator's experience is the magnitude of the perpendicular component. The larger 
the \emph{perpendicular force} becomes, the more difficult it is for the operator to continue to
persue his\slash her intended direction. This is, however, not always an unwanted feature. Although 
it seems to be merely impedimental, important parts of the response are encoded in the perpendicular 
force component. On the fly, it directs the operator to alternative reaction valleys (minimum energy
paths) that are different from the intended one. This feature is welcome as it works against a 
potential bias of the operator and thus helps to detect unexpected pathways. 

Another more technical issue is what we may call the problem of \emph{evasive adaptation}. It arises 
solely in manipulations against a barrier. The resistance of the operator to a force 
response of the system on the manipulated fragment generates an opposing force on all other atoms. 
In the case of a simultaneously running energy optimization, the molecular system will adapt to the new 
constraint (that is the fixed atoms) by moving the remaining atoms until the forces on them vanish. 
If the manipulation is carried out sufficiently slowly so that the system can always adapt, any 
attempt to overcome a barrier will fail.

Evasive adaptation is a characteristic of the energy optimization procedure that is constantly 
trying to reach the closest minimum on the PES. A typical example is an addition reaction with a 
barrier. If the operator pushes a fragment onto another one, the other fragment will simply be 
pushed away. However, evasive adaptations can also be a useful feature. Assume, for instance, that 
an operator grabs a fragment by picking one atom to move it in a neutral manipulation. Depending on 
the movement's velocity two different events can occur. Either the fragment is torn apart by 
breaking the weakest bond or the whole fragment just follows the moved atom. The latter can be 
enforced, if parts of the system are treated as rigid bodies with no internal degrees of freedom but 
with free overall translation and rotation. 

Such artificial constraints mimic a second input device that may keep parts of the system in a 
suitable position and orientation for the exploration. The energy minimization procedure acts then 
only on a subspace of the configuration space. The constraints can, of course, also be directly 
applied by additional haptic devices or other input devices, either operated by the same operator or 
in a collaborative manner by more than one person. The operator(s) then need(s) to decide when to 
relax the constraints to avoid exploring artificial reaction paths. The algorithm can provide 
assistance by rendering (visually or haptically by an additional device) the forces on the 
constrained fragments. Large forces on single atoms are a clear indication that the corresponding 
fragment would adapt to a structural manipulation but is hindered by the constraints.

A more general approach is to allow a continuous change between the two scenarios solely based on 
the speed of the manipulation. For this the quantum chemical structure optimization procedure needs 
to be adjusted such that a rather slow manipulation of a fragment leads to a translation of the 
whole fragment and a fast manipulation leads to bond breaking. Then, the operator can intuitively 
control the behavior without the need of additional devices or other artificial constraints. Fast 
manipulations with the intention to overcome barriers would then correspond to nonadiabatic 
manipulations, whereas slow manipulations with an instantaneous structure adaptation would be 
classified as adiabatic. 

If the computational demands of an electronic structure and force calculation (and force evaluation) 
set an upper limit to the optimization speed, the manipulation has to be slowed down until the 
optimization is faster than the manipulation. This can be easily achieved by modifying the 
transformation of the haptic pointer position from the world into the molecular frame. 

\subsection{Global Reactivity}

\begin{figure}[h!]
    \centering
    \includegraphics[width=0.80\textwidth]{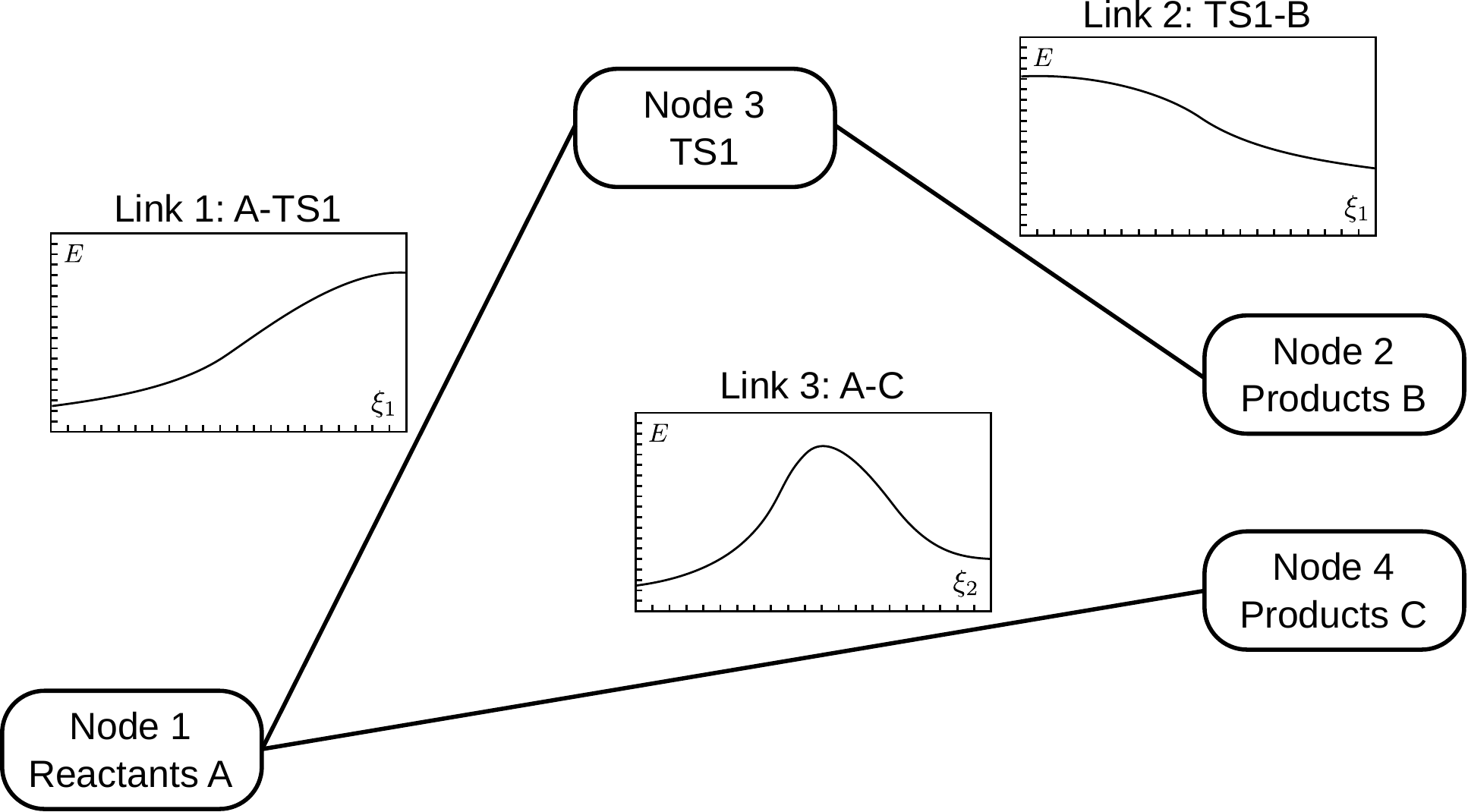}
    \caption{Elements of a reaction network representing the results of a global reactivity 
    exploration. The network shows two reaction paths starting from reactants at Node 1 moving on to 
    the products Nodes 2 and 4. The reaction path from reactants A to products B has already been 
    analyzed as indicated by an additional node (Node 3) corresponding to the transition state TS1. 
    Link 3 shows a link before a closer examination, which would be followed by an extraction of a 
    transition state node and a split of the link. The nodes are arranged according to their 
    energies. Note that Link 1 and 2 belong to the same reaction coordinate ($\xi_1$), whereas Link 
    3 has a different one ($\xi_2$)} \label{fig:globalreactivity}
\end{figure}

Exploring global reactivity requires to organize and manage the data obtained during an interactive 
reactivity exploration and to present it in such a way that the operator quickly obtains an 
overview. Graphs are a tool to store provisional results of local reactivity studies (see Figure 
\ref{fig:globalreactivity}). The stored structures are the starting points for further explorations 
and for the refinement of existing ones. The more structures the operator collects the more the 
network graph also serves as a tool to organize the results. The operator should be able to change 
the level of detail for the whole graph, for sub-graphs or even for single pairs of nodes. 

In general, the shear number of possible reactions and rearrangements in large assemblies makes it 
impossible to obtain a simple picture of the system's entire reactivity. One may aim for a 
representation of an essential subset of chemical transformations represented as a network of local 
minima and transition states connected by (elementary) reaction paths. These networks may harbor 
circles as in catalytic pathways as well as tree-like structures.

Collections of molecular structures forming a discretized path through configuration space connect 
the nodes of the reaction network as links. Optionally, energies and other molecular properties can 
be stored for each structure of the link, e.g., to provide an energy profile (see insets in Figure 
\ref{fig:globalreactivity}). Clearly, convenient use of the graph implies that data retrieval (e.g.\
of structures) is achieved by a single mouse click. 

Nodes may be stored automatically when the structure gradient drops below a certain threshold or 
upon active selection of a structure by the operator. Once a sufficient number of nodes and links 
is recorded, the operator can analyze the paths in the links. Interesting structures might be marked 
as entry points for a later exploration. 

\section{Implementation in the {\sc Samson} framework}
\label{sec:implementation}

The above outlined features necessary for an interactive study of chemical reactivity have been
implemented in a couple of programs (called Apps) that can be loaded into the {\sc Samson} environment. 
Since the graphical user interface (GUI) of the {\sc Samson} program is based on the Qt library \cite{qt}, 
Qt classes are heavily used in our Apps, too. For details on the Qt classes employed we refer the 
reader to the official Qt documentation \cite{qt}. In the following, we use the term GUI 
collectively for all windows, panels, buttons etc.\ that are part of our Apps. 

The density-functional tight-binding (DFTB) App delivers the energies and forces needed for the
exploration of molecular reactivity in real time. It implements the standard non-self-consistent 
variant of DFTB. The functionality for local reactivity studies is split into two separate Apps 
called Local Reactivity and Phantom Direct. All features concerning the energy minimization and the 
visual display options are collected in the Local Reactivity App, whereas all features to influence 
the force rendering and the input behavior of the haptic pointer device are combined in the Phantom
Direct App. The global reactivity App is called Global Reactivity Monitoring. It allows the operator 
to store, organize, and visualize reaction networks, which are the result of reactivity studies as 
described in the preceding sections.

\subsection{DFTB App}

In this section, we assess the capabilities of DFTB for interactive reactivity studies. The standard 
non-self-consistent DFTB method\cite{porezag1995, seifert1996} is a highly-parametrized 
tight-binding method, where all parameters are obtained from DFT calculations and not by fitting 
them to experimental data.

The DFTB method is based on a Taylor expansion of the DFT energy around a reference electronic 
density\cite{elstner2014}, which is a superposition of the atomic densities. It neglects second and 
higher-order terms and its energy reads
\begin{align}\label{eq:dftbenergy}
 E^\mathrm{DFTB} = E_\mathrm{rep} + \sum_i^\mathrm{occ} \varepsilon_i .
\end{align}

The repulsion energy $E_\mathrm{rep}$ is calculated by summing up the pairwise repulsion 
interactions of the atoms, 
\begin{align}
 E_\mathrm{rep} = \sum_{I < J} V_{IJ}^\mathrm{rep} 
\end{align}
The functions $V_{IJ}^\mathrm{rep}(R_{IJ})$ are piecewise constructed from analytical functions of 
the distance $R_{IJ}$ between the atoms $I$ and $J$. They consist of an exponential function at 
short distances, of third-order polynomials for mid-range distances, and of a fifth-order polynomial 
that smoothly decreases the repulsion to zero at a given cutoff distance. The coefficients 
describing these functions are available in parameter sets\cite{elstner1998, gaus2013, gaus2014b}.

The second term on the right hand side of Eq.~(\ref{eq:dftbenergy}) is the binding energy based on 
an linear combination of atomic orbitals taking into account only the valence orbitals. It leads to 
a generalized eigenvalue problem,
\begin{align}
 \boldsymbol{H}^0\boldsymbol{C} = \boldsymbol{S}\boldsymbol{C}\boldsymbol{\varepsilon},
\label{eq:GEP}
\end{align}
delivering the eigenvector matrix $\boldsymbol{C}$ and the diagonal matrix 
$\boldsymbol{\varepsilon}$ containing the single-particle energies $\varepsilon_i$. For atomic 
orbitals $\mu$ and $\nu$ belonging to the same atom, the Hamiltonian matrix elements $H_{\mu \nu}^0$ 
are given by the atomic orbital energies for $\mu = \nu$ and zero for $\mu \neq \nu$. Similarly, 
the overlap matrix elements for atomic orbitals of the same atom are given by the Kronecker delta. 
For atomic orbitals belonging to different atoms, the matrix elements $H_{\mu \nu}^0$ and 
$S_{\mu \nu}$ are calculated by performing Slater-Koster transformations\cite{slater1954} of 
integrals obtained by an interpolation of values tabulated for atom pairs at given distances. These 
tables are available in the DFTB parameter sets\cite{elstner1998, gaus2013, gaus2014b}.

The force $\boldsymbol{f}_A$ acting on an atom $A$ is given by the negative gradient of the energy 
$E^\mathrm{DFTB}$ with respect to the corresponding nuclear coordinate\cite{elstner1998}, 
\begin{align}
 \boldsymbol{f}_A = - \frac{\partial E^\mathrm{DFTB}}{\partial{\boldsymbol{R}_A}}
  = - \frac{\partial E_\mathrm{rep}}{\partial{\boldsymbol{R}_A}}
    - \sum_{\mu \nu} \left(  P_{\mu \nu} \frac{\partial H_{\mu \nu}^0}{\partial{\boldsymbol{R}_A}}
          - W_{\mu \nu} \frac{\partial S_{\mu \nu}}{\partial{\boldsymbol{R}_A}} \right),
\end{align}
where $\boldsymbol{P} = \{ P_{\mu \nu} \}$ is the density matrix and $\boldsymbol{W} = \{ W_{\mu 
\nu} \}$ is the energy-weighted density matrix. As the DFTB method does not involve any integral 
evaluation and Eq.~(\ref{eq:GEP}) does not require an iterative solution, the method is 
computationally inexpensive, which is a considerable advantage for interactive reactivity 
exploration (if no specialized hardware and parallelism is exploited).

In our implementation, all matrix elements for a given atom pair are calculated in one shot in order 
to avoid redundant operations and minimize the execution time. The implementation of the 
interpolation for the matrix elements was inspired by the DFTB+ program\cite{aradi2007}. For 
interatomic distances covered by the tabulated values, an eigth-order interpolation scheme following 
Neville's algorithm\cite{numericalRecipes} has been chosen. The matrix elements for interatomic 
distances larger than the tabulated values are calculated by applying a fifth-order extrapolation 
that smoothly decreases them to zero. The coefficients for the extrapolation are determined during 
the initialization of the method. The derivatives of the Hamiltonian and overlap matrix elements are 
obtained by numerical differentiation of the tabulated values and application of the product and 
chain rules for the Slater-Koster transformations. For the solution of the generalized eigenvalue 
problem in Eq.~(\ref{eq:GEP}), the corresponding function of the Intel$^\text{\textregistered}$ 
Math Kernel Library\cite{mkl} has been employed. To dynamically adapt the thickness of bonds 
rendered by {\sc Samson}, the Mayer bond-order matrix\cite{mayer1983} is calculated from the density and 
overlap matrices. We applied the MIO\cite{elstner1998} and 3OB\cite{gaus2013, gaus2014b} parameters, 
designed for organic molecules containing H, C, N, O, P and S atoms. Although these sets were 
originally developed for the self-consistent-charge DFTB variants DFTB2\cite{elstner1998} and 
DFTB3\cite{gaus2011}, they yield reasonable molecular structures when we employed them in the 
non-self-consistent DFTB method. Table \ref{table:timings} shows the execution times for the energy 
and gradient calculation of a set of test molecules indicating the feasibility of (unparallelized) 
DFTB for real-time quantum chemical studies on molecules of moderate size. 

\begin{table}[!htb]
\caption{Execution times for the DFTB energy and gradient calculation of different molecular systems.
  The timings were performed on an Intel$^\text{\textregistered}$ Core$^\text{\texttrademark}$ i7 
  (3.20~GHz) workstation.\label{table:timings}}
\begin{tabular}{l c c c}
\toprule
 System & Molecular formula & Number of orbitals & Execution time \\
\midrule
 Methane & CH$_4$ & 8 & \textless{} 1 ms \\
 Diels-Alder & C$_2$H$_4$ + C$_4$H$_6$ & 34 & \textless{} 1 ms \\
 Parathion &  C$_{10}$H$_{14}$NO$_{5}$PS & 96 & 3 ms \\

 Oleic acid & C$_{18}$H$_{34}$O$_{2}$ & 114 & 4 ms \\
 Vitamin A & C$_{20}$H$_{30}$O & 114 & 4 ms \\ 
 Endiandric acid C & C$_{23}$H$_{24}$O$_{2}$ & 124 & 5 ms \\
 Folic acid & C$_{19}$H$_{19}$N$_7$O$_{6}$ & 147 & 7 ms \\
 Cholesterol & C$_{27}$H$_{46}$O & 152 & 8 ms \\
 Fexofenadine & C$_{32}$H$_{39}$NO$_{4}$ & 187 & 11 ms \\
 Quinovic acid & C$_{30}$H$_{46}$O$_{5}$ & 186 & 12 ms \\
 Taxol & C$_{47}$H$_{51}$NO$_{14}$ & 206 & 15 ms \\ 
 NADH & C$_{21}$H$_{29}$N$_7$O$_{14}$P$_2$ & 215 & 15 ms \\
 Fullerene & C$_{60}$ & 240 & 20 ms \\
 $\alpha$-Cyclodextrin & C$_{36}$H$_{60}$O$_{30}$ & 324 & 31 ms \\ 
 Vasopressin & C$_{46}$H$_{65}$N$_{13}$O$_{12}$S$_{2}$ & 375 & 46 ms \\
 Valinomycin & C$_{54}$H$_{90}$N$_6$O$_{18}$ & 402 & 54 ms \\
 Tannic acid & C$_{76}$H$_{52}$O$_{46}$ & 540 & 95 ms \\ 
\bottomrule
\end{tabular}
\end{table} 

We assess the reliability of DFTB by comparing it to standard quantum chemical methods. The 
reference data were obtained with the B3LYP exchange-correlation functional\cite{becke1993,
lee1988,stephens1994} in combination with the aug-cc-pVTZ\cite{dunning1989} and def2-TZVP
\cite{weigend2005} basis sets as well as with the coupled-cluster singles, doubles, and perturbatively
treated triples, CCSD(T), approach\cite{raghavachari1989} in combination
with the Dunning aug-cc-pVTZ basis set in spin-unrestricted calculations.
The results were also compared to the ASED-MO method\cite{anderson1994}, which is available in {\sc Samson}\cite{bosson2012} and
which was applied to demonstrate structure optimization in real time. For ASED-MO, no explicityl 
optimized parameters for organic compounds are available. The parameters of Ref.~\cite{diaz2001} 
were employed in combination with a STO-2G basis\cite{hehre1969} for the calculation of the overlap 
matrix. We note that studying bond breaking processes with standard DFT methods require the breaking
of the spin symmetry\cite{jacob2012}.

\begin{figure}[!htb]
\begin{center}
 \includegraphics[width=\textwidth]{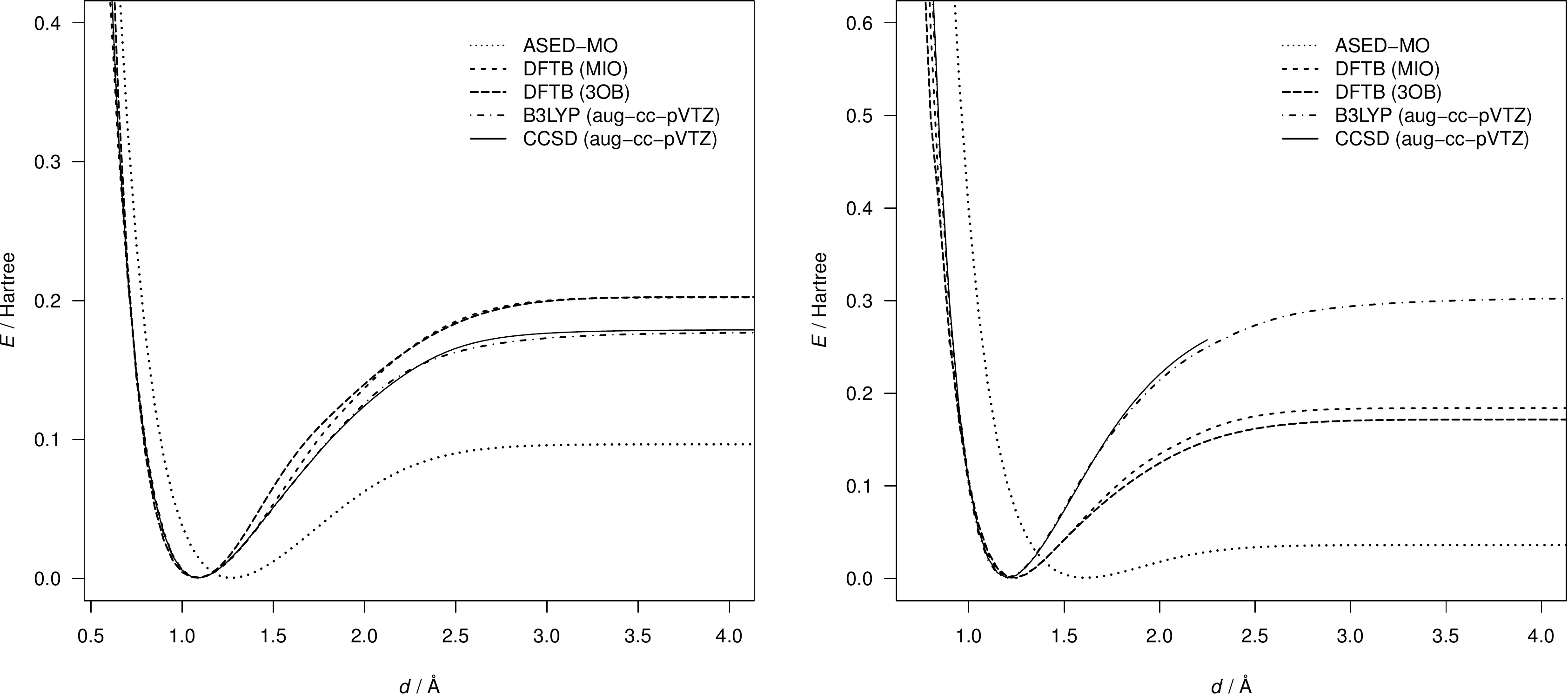}
\caption{Energy profile for the dissociation of a C-H bond in methane (left) and the C=O bond in 
formaldehyde (right) as a function of the bond length $d$. At every distance, the remaining 
coordinates of the system were allowed to relax in order to minimize the energy. The DFTB profiles 
calculated with the MIO and 3OB parameters for the C-H bond cannot be distinguished except for bond 
lengths around 1.7~\AA{}. The reference calculations were carried out with the {\sc Turbomole}
program package\cite{ahlrichs1989} (version 6.4). Special care has been taken that a truly
unrestricted (i.e. broken-spin-symmetry) Kohn--Sham determinant has been converged asymptotically.
The energy scale is chosen to be zero for the lowest energy.\label{fig:CHCObonds}}
\end{center}
\end{figure}

\begin{figure}[!htb]
\begin{center}
 \includegraphics[width=\textwidth]{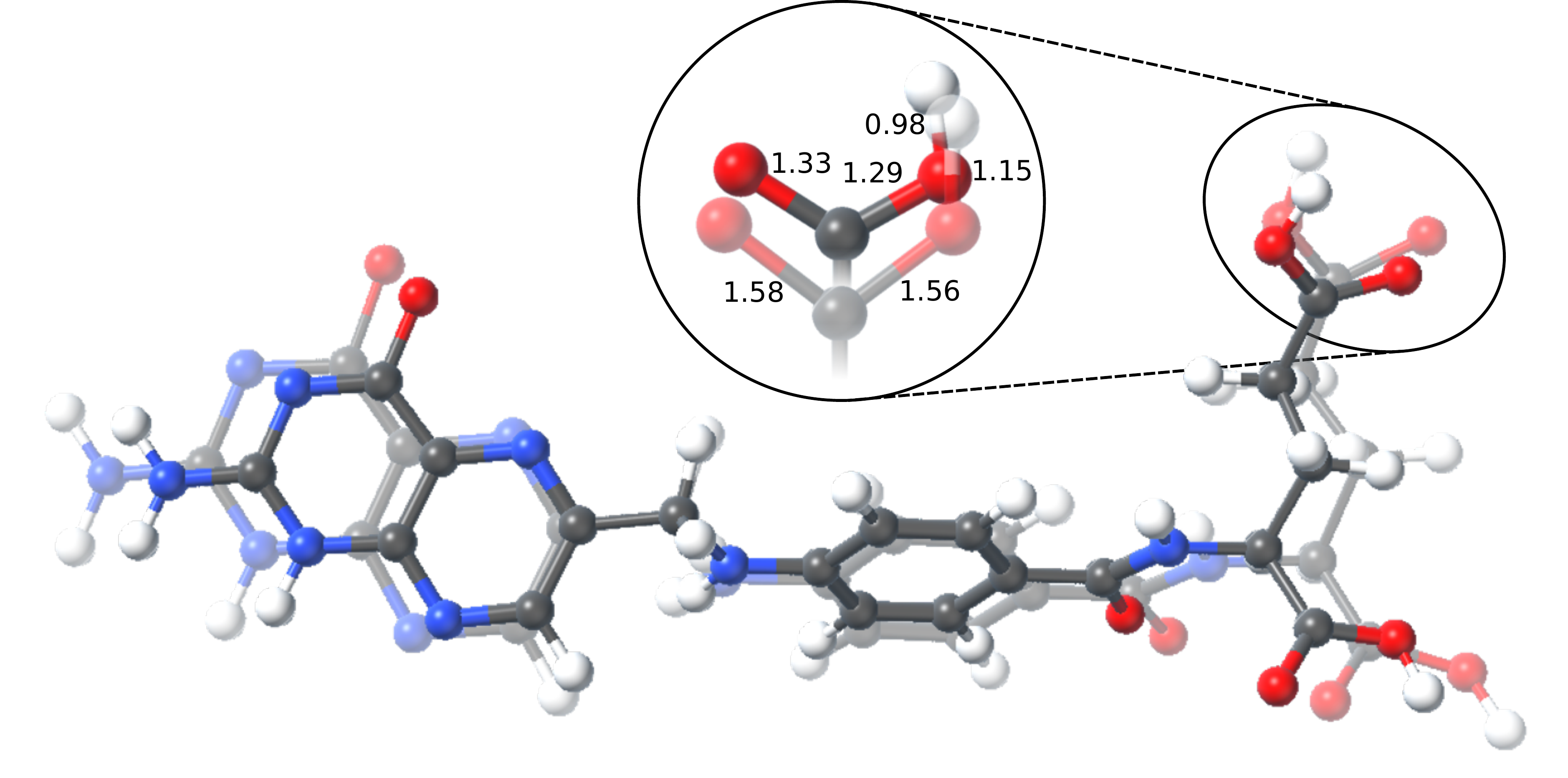}
\caption{Equilibrium structures of folic acid predicted by the DFTB method (opaque) and the ASED-MO 
method (partially transparent). The bond lengths in one of the carboxyl groups are indicated 
in~\AA{} in the inset. The bond thickness is an indicator for the calculated Mayer bond order. 
\label{fig:folicAcid}}
\end{center}
\end{figure}

Figure \ref{fig:CHCObonds} displays energy profiles for the breaking of a C-H bond in methane and of 
the C=O bond in formaldehyde. The energy profile produced by the DFTB method for the C-H bond is in 
good agreement with the reference DFT calculation. The equilibrium bond length does also not deviate 
much from the reference value for the C=O bond, but the predicted energy well is too wide. For the 
haptic exploration of molecular systems, this means that the user feels a too weak force and 
therefore an inaccurate representation of the stiffness of the carbonyl bond. This inability of the 
DFTB method to describe polarized systems correctly is well known, since the non-self-consistent 
character of the method does not allow for charge transfer. Moreover, the DFTB energy profiles 
exhibits too small dissociation energies. This is not a major drawback as the haptic exploration of 
the molecular system requires only a qualitatively correct potential energy surface\cite{haag2014a}. 
Figure~\ref{fig:CHCObonds} also shows that the ASED-MO method with the chosen parameters 
overestimates equilibrium bond lengths. The differences between the molecular structures obtained 
with DFTB and ASED-MO are illustrated at the example of folic acid in Figure~\ref{fig:folicAcid}. 
The energy wells predicted by the ASED-MO method are wider and shallower than their DFTB 
counterparts, which implies easier bond dissociations during the haptic exploration. As every 
parametrized method, also ASED-MO could be adapted to yield improved bond lengths at the expense of 
transferability to other systems.

\begin{figure}[!htb]
\begin{center}
 \includegraphics[width=0.6\textwidth]{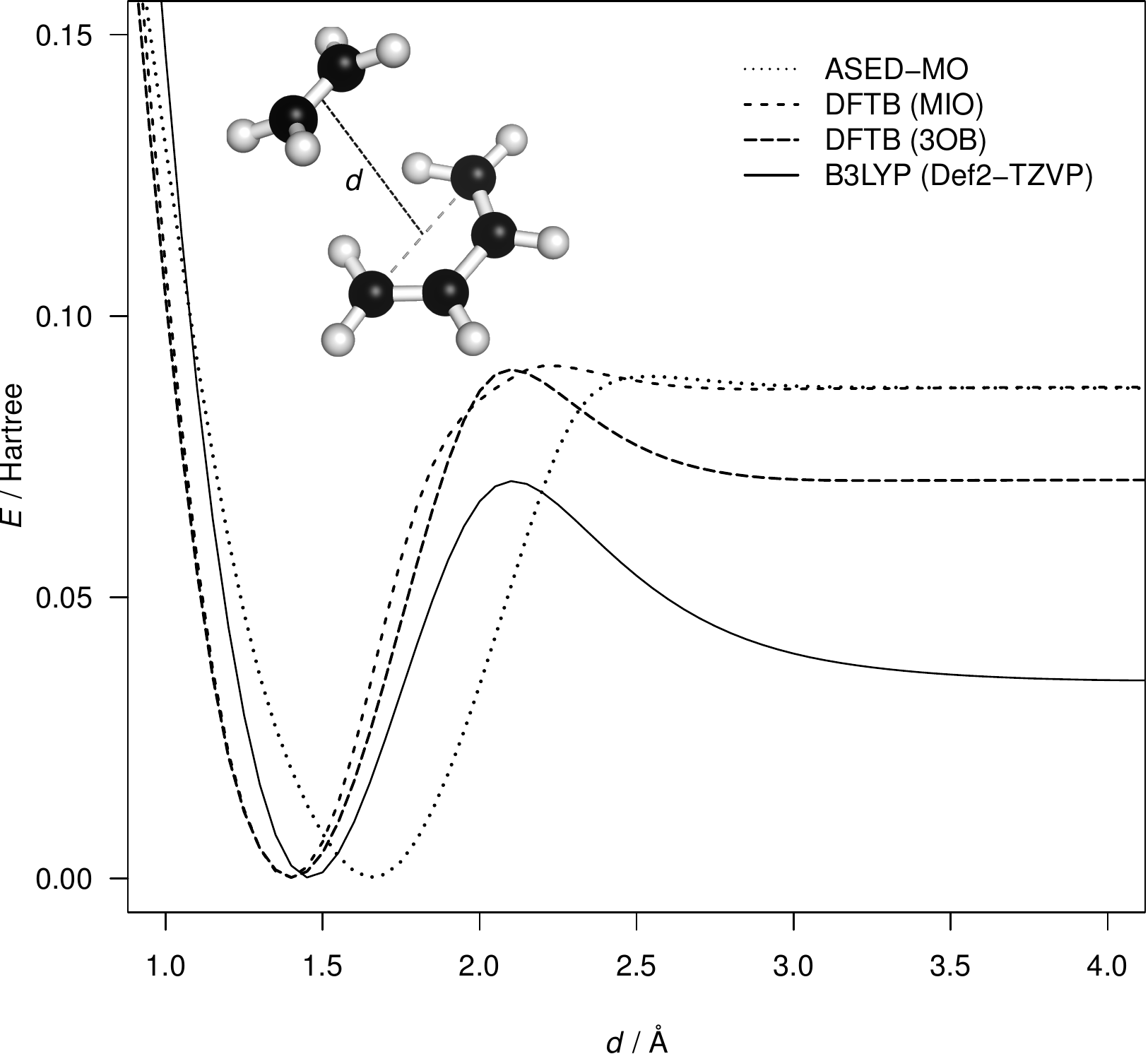}
\caption{Energy profile for the Diels-Alder reaction between ethene and butadiene. 
The reaction coordinate is the distance $d$ between the two lines defined by the two carbon 
atoms of ethene and the two terminal carbon atoms of butadiene, drawn above as a dashed line. At 
each distance, the structure was allowed to relax while requiring the two lines to remain parallel.
The reference calculation was carried out with the Gaussian 09 program package\cite{gaussian09} (rev.\ d.01).
The energy scale is shifted to be zero for the lowest energy. \label{fig:diels-alder}}
\end{center}
\end{figure}

Figure \ref{fig:diels-alder} shows an energy profile of the Diels-Alder reaction of butadiene and 
ethene obtained with different methods. The reaction coordinate has been chosen to be the distance
between the two parallel lines defined by the two carbon atoms of ethene and the two terminal carbon 
atoms of butadiene. The DFTB method in combination with the 3OB parameters produces a qualitatively 
correct energy profile, also compared to previous studies of the reaction\cite{sakai2000}. 
Furthermore, the positions of the energy minimum and maximum are in a good agreement with the 
B3LYP/def2-TZVP results. The MIO parameters for DFTB and the ASED-MO method fail to reproduce a 
significant barrier for this reaction pathway. For ASED-MO, a parametrization optimized for this 
special system might be found that results in a better representation of the barrier. 

\begin{figure}[!htb]
\begin{center}
 \includegraphics[width=\textwidth]{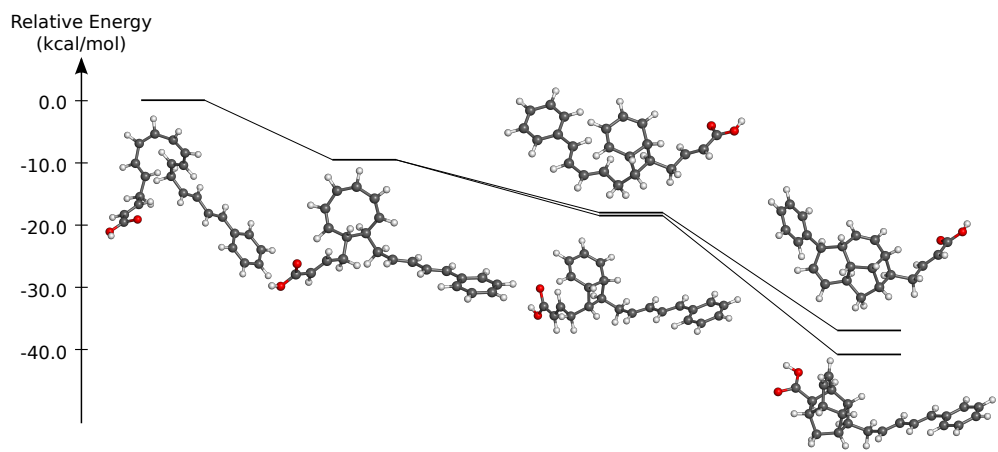}
\caption{Structures and energies obtained during the exploration of the biosynthesis pathway of 
endiandric acid B (upper pathway) and C (lower pathway). The first step is an electrocyclic reaction 
leading to a common intermediate, that undergoes another electrocyclic reaction in the second step, 
producing one of two possible structures. The third step is in both cases a Diels-Alder reaction.}
\label{fig:endiandricAcid}
\end{center}
\end{figure}

The reactivity of several molecular systems was investigated with DFTB in the {\sc Samson} framework. The 
haptic exploration of systems containing less than about 120 atoms proceeds very smoothly. In 
particular, we observed that the DFTB method allows a remarkably reliable exploration of pericyclic 
reactions. For example, the biosynthesis of the endiandric acids B and C postulated by Bandaranayake 
et al.\cite{bandaranayake1980}, which features a cascade of three pericyclic reactions, could be 
reproduced correctly using the haptic device. The structures of the reactant, intermediates and 
products as well as the corresponding energies are given in Figure \ref{fig:endiandricAcid}.

\subsection{Local Reactivity App}

\begin{figure}[!htb]
  \centering
  \includegraphics[width=\textwidth]{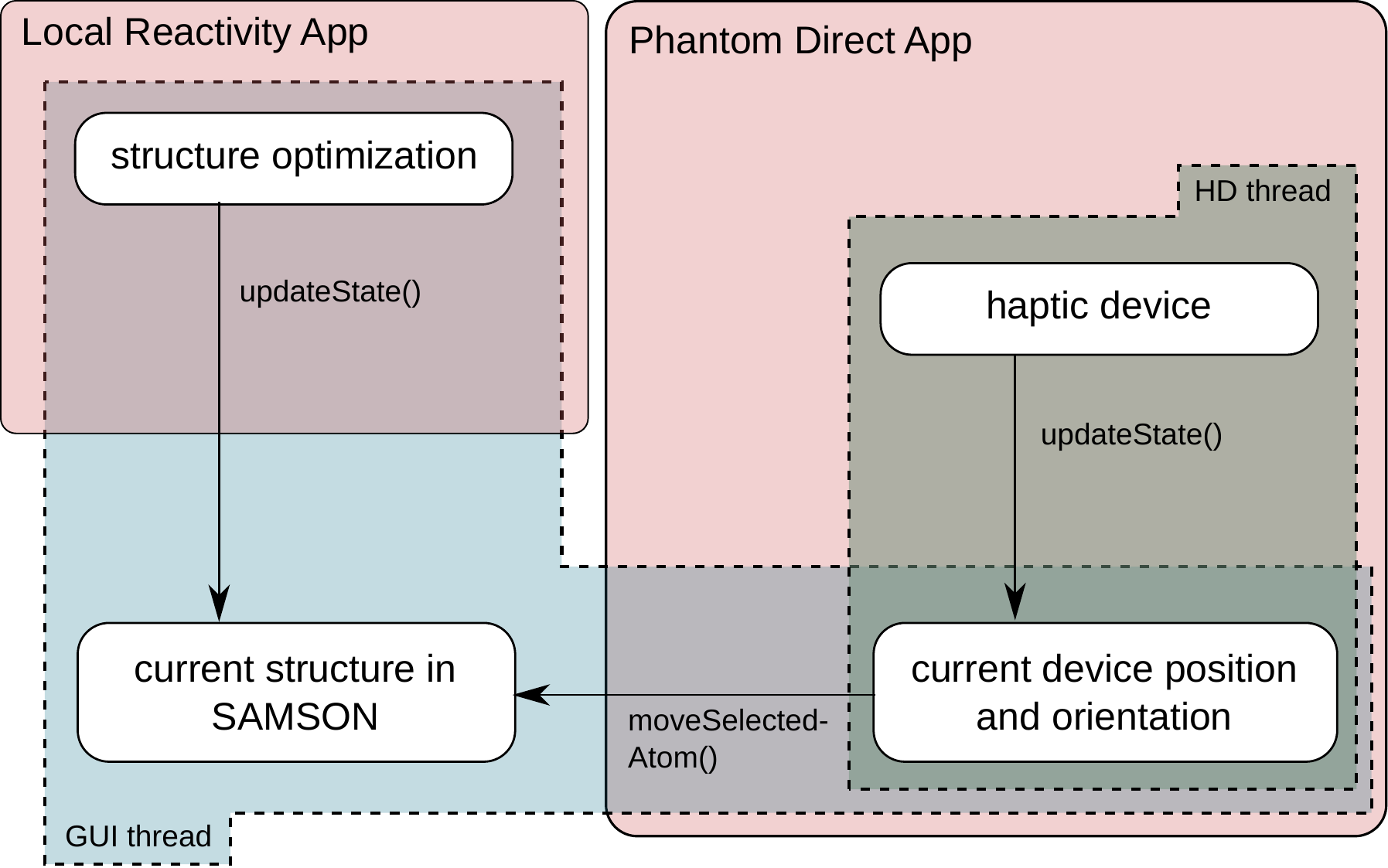}
  \caption{Schematic picture to illustrate the interplay between the GUI thread and the
haptic device (HD) thread and the implementation of the loops in the two different local reactivity 
Apps. The different loops are represented by arrows with the name of the function attached which 
implements their functionality.
  \label{fig:threadsandapps}}
\end{figure}

The implementation of the local reactivity Apps focuses on the interplay of continuously running 
loops, whereas the global reactivity App has its focus on the data model developed to store reaction 
networks. Most of the loops are implemented as functions continuously executed by timers provided by 
Qt and, thus, are all part of the GUI thread. This implies that these tasks are executed 
sequentially, which avoids the difficulties that come along with the parallel execution of code. The 
only exception are tasks which need to communicate with the haptic device. They are executed in a 
separate thread maintained by the device driver. If the loops, which are part of the GUI thread, 
become too slow (e.g., for large molecules or more accurate and thus computationally expensive 
electronic structure calculations), additional threads can be started to allow a parallel execution 
of the most time demanding tasks. Then, however, the involved data objects need to be made 
thread-safe. Figure \ref{fig:threadsandapps} provides an overview over the different loops, threads 
and their distribution over the two Apps responsible for local reactivity (Local Reactivity App and 
Phantom Direct App).

The next step is to choose the algorithm for the energy minimization procedure. Currently we have 
implemented a steepest descent and a conjugate gradient algorithm. Although the latter is generally 
more stable and converges faster than the former, in most cases steepest descent will be the method 
of choice. It has the advantage of always converging to a minimum and of being computationally cheap 
and therefore fast. More elaborate algorithms will rarely be beneficial in our design of virtual 
environments (VEs), since during an exploration the configurations to be minimized are always very 
close to a minimum. However, the modular structure of {\sc Samson} makes it easy to load other and 
more complex algorithms such as adaptive integrators\cite{rossi2007}. In our steepest descent 
implementation, the step size and also the number of steps per optimization cycle can be adjusted in 
a GUI panel that pops up when the algorithm has been selected (see Figure 
\ref{fig:localreactivitygui} in the appendix). In our studies conducted so far, a step size of 
$0.02$ and $5$ steps per cycle were sensible start values. Depending on the shape of the PES the 
step size might need to be adjusted in order to prevent oscillations around the minimum. If it turns 
out during an exploration that the parameters are not suitable anymore, the parameters as well as 
the method can be changed on-the-fly.

\subsection{Phantom Direct App}

With the Phantom Direct App the operator controls the force rendering and is able to adjust all 
possible scalings that are applied when the molecular force calculated by the electronic-structure
method is transformed to the force actually rendered by the device. 

The force processing consists of scaling and damping the molecular force to eventually produce the
haptic force passed to the device, 
\begin{align}
    \bm{f}_\mathrm{out} = c(\bm{f}_\mathrm{in}) \, \bm{f}_\mathrm{in} \,\, ,
\end{align}
where $c$ may be a function of the force but is subjected to certain conditions. It has to be a 
strictly positive function of the input force $\bm{f}_\mathrm{in}$ and is not allowed to have any 
zero crossings (apart from the one at the origin), since they would introduce artificial extremal 
points on the surface\cite{haag2014a}. The scaling function chosen for our implementation takes care
of the necessary unit conversion and the adaptation of the force range to the range of the haptic 
device. For details we refer the reader to the corresponding section in the appendix.

A decomposition of the force into components parallel and perpendicular to the intended direction
is necessary to scale them separately. As the intended direction, the normalized and damped velocity 
vector of the end-effector (tip of the ``pen'' of the haptic device) are taken (the velocity is 
damped in the same way as the force). The decomposition and scaling will only be performed if the 
length of the velocity vector is above a certain threshold. The decomposition is carried out as 
stated in Eqs.~(\ref{eq:forcedecomp1}) and (\ref{eq:forcedecomp2}). The scaling is then achieved by 
summing up the components
\begin{align}\label{eq:forcedecomp}
  \bm{f}_\mathrm{h} = c_\perp \bm{f}_\perp + c_\parallel \bm{f}_\parallel \quad
\textnormal{with} \quad c_\perp, c_\parallel \geq 0 \quad \textnormal{and} \quad c_\perp +
c_\parallel > 0 \,\, ,
\end{align}
where the coefficients $c_\perp$ and $c_\parallel$ determine the scaling. Because of the conditions 
for the coefficients, the force direction cannot be rotated by more than 90 degrees, thus assuring 
that an attractive force never becomes repulsive and vice versa. The condition on the sum (last 
condition in Eq.~(\ref{eq:forcedecomp})) prevents an introduction of artificial zero-crossings. 

The molecular input force $\bm{f}_\mathrm{m}$ in the simulation loop is updated at a lower frequency 
than the one in the device loop, since the quantum chemical calculation of the former can hardly be 
achieved at a kilohertz rate. Thus, we have to apply a damping scheme before the force is passed to the device. This is achieved by calculating
the force to be rendered as a superposition 
\begin{align}\label{eq:damping}
  \bm{f}^\mathrm{(h)}_\mathrm{m} = c_\mathrm{p} \bm{f}^\mathrm{(p)}_\mathrm{m} + c_\mathrm{n} \bm{f}^\mathrm{(n)}_\mathrm{m} \quad
  \textnormal{with} \,\, c_\mathrm{p} + c_\mathrm{n} = 1
\end{align}
of the previously rendered force $\bm{f}^\mathrm{(p)}_\mathrm{m}$ and the new (scaled) force
$\bm{f}^\mathrm{(n)}_\mathrm{m}$. In our applications, we usually choose $c_\mathrm{n} = 0.05$ for 
the damping. The damping not only smoothes the force experience but also prevents fast oscillations 
when the haptic pointer is trapped in the valley of a steep minimum. 

\begin{figure}[!htb]
  \centering
  \includegraphics[width=\textwidth]{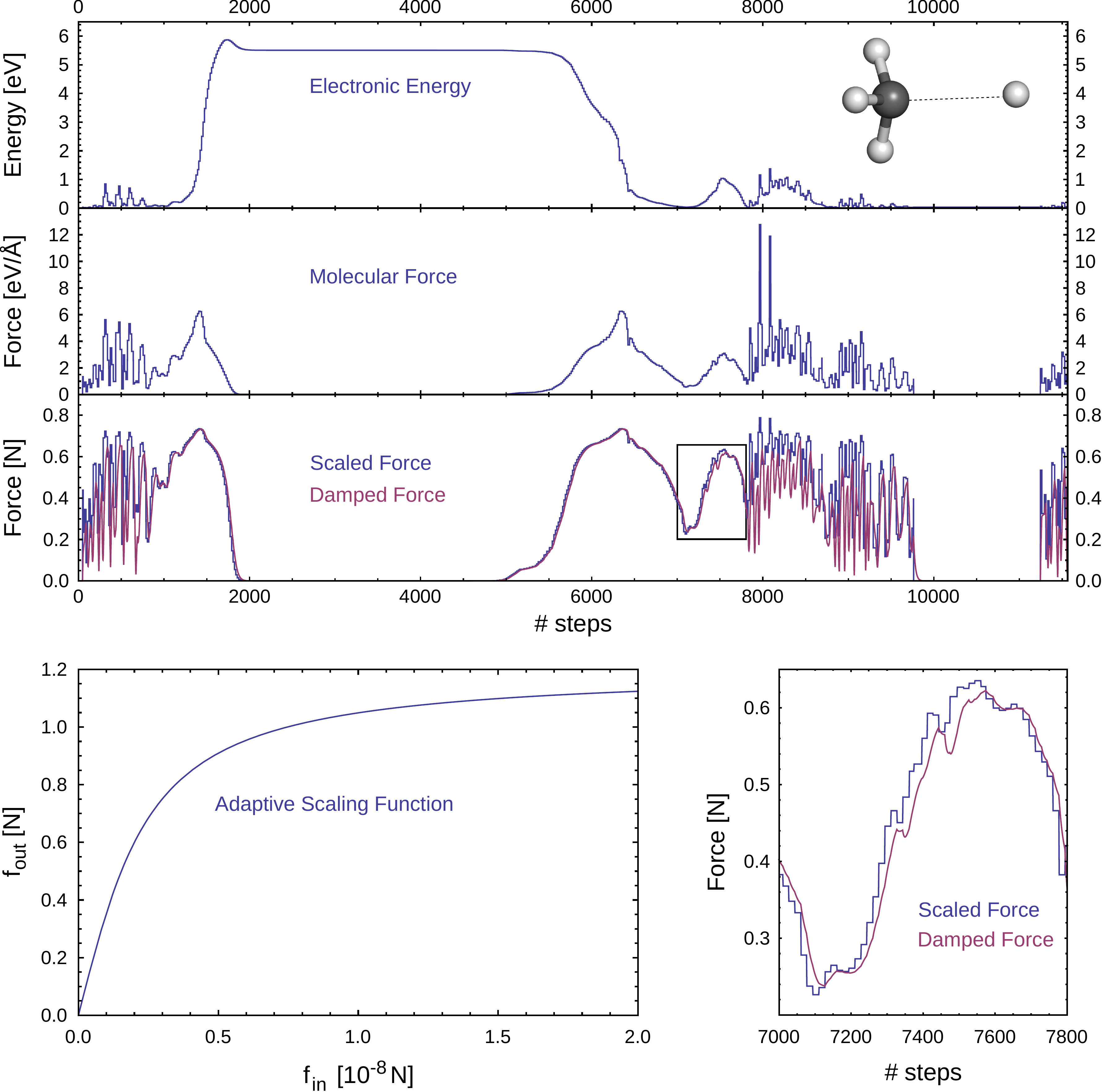}
  \caption{Plots of the electronic energy and the total force magnitudes during exploration of the 
binding energy of methane (abstraction of a hydrogen atom) employing our non-self-consistent 
DFTB implementation. The energy and forces have been recorded at every iteration ($\equiv 1 
\mathrm{step}$) of the device loop. The adaptive scaling was performed as in 
Eq.~(\ref{eq:adaptivescaling}) with $m = 0.5\times10^{9}\,\mathrm{N}^{-1}$, $f_\mathrm{max} = 
1.2\,\mathrm{N}$. The adaptive scaling function is plotted as a function of the force in Newton,
whereas the molecular force is given in $\mathrm{eV}/\text{\AA}$. The adaptively scaled and 
damped (as in Eq.~(\ref{eq:damping}) with $c_\mathrm{n} = 0.05$) force curves are plotted on top of 
each other. The bottom right plot magnifies the indicated part in the plot of the scaled and damped 
force to show the effect of the damping.\label{fig:scalingsDampingsPlot}}
\end{figure}

The effects of the scaling and damping are illustrated in Figure \ref{fig:scalingsDampingsPlot} at 
the example of the abstraction of a hydrogen atom from a methane molecule. The energy is shifted to 
be zero for the equilibrium structure of methane. The forces and energies were recorded at every 
pass of the device loop resulting in $1000$ data points (steps) per second. The plateau occurring 
roughly between step $2000$ and $5500$ corresponds to the manipulated hydrogen atom being at the 
dissociation limit. The smaller peaks between step $0$ and step $1000$ and between step $7000$ and 
step $10000$ are due to distortions from the minimum structure by pulling slowly to allow the 
relaxation procedure to adapt the positions of the remaining atoms. By this, the operator was 
actually moving the whole methane molecule. The force curves show the forces calculated by the 
electronic structure method and illustrate why the adaptive scaling is needed to keep the forces 
within the range suitable for the device. The very high force peaks are reduced while preserving the 
subtle differences when the calculated force is comparatively small. From the bottom right plot in 
Figure \ref{fig:scalingsDampingsPlot} one can clearly see that the simulation loop and the device 
loop are running at different frequency yielding a step function for the rendered force. This is 
smoothed out by the damping function. The magnified part of the plot shows a possible drawback of 
the damping. Due to the delay introduced by the damping, force changes that occur at a too high 
frequency vanish. A too strong damping may lead to artificially flat surfaces by hiding subtle 
details, but in general, the elimination of high-frequency force changes is wanted. However, in the 
example shown in Figure \ref{fig:scalingsDampingsPlot} the damping does not hide the surface 
features (force changes) that are characteristic for the reactivity, since they occur at a much 
lower frequency. 

\subsection{Global Reactivity Monitoring App}\label{sec:globalreactivity_app}

The Global Reactivity Monitoring App serves as a tool to collect, organize, and evaluate the results 
of a reactivity study. Since stable intermediates and transition state structures are the main 
result of a reactivity study, the App records and presents them in a graph of the network. Details 
on the implementation or the data model developed are given in the appendix. A link containing a 
reaction path can be created by recording a path. This requires that the current molecular structure 
displayed in {\sc Samson} corresponds to one node (i.e., one node needs to be active) which will then be
the starting node for the link. When the recording is active, the App will react to every structural 
signal emitted by {\sc Samson} (every change of an atom position during an exploration leads to a new 
structural signal) by calculating the RMSD between the current structure and the last recorded 
structure and adding the new structure as a node to a buffer if the RMSD is above a threshold. When 
the recording is stopped, a new link is created and the elements of the buffer are moved into the
link object. The starting node of the link points to the node that was active at the beginning and 
the end node is the last node added to the buffer. Both, the start and the end node, are not part of 
the path, hence are not part of the link. During the recording the energy profile of the reaction 
path is plotted in the lower panel of the App. At every click on a link that stores a reaction path 
a window pops up with a more detailed plot of the reaction path, on which the operator can analyze 
the path by zooming in and out. 

New nodes and links can also be added by splitting an existing link with a stored reaction path. The 
operator can choose a link and specify at which position (which element of the path) the link should 
be split. The node corresponding to this position will then be promoted to be a top-level item and 
will be equipped with a corresponding graphics item. Two new links will be created where the new 
node is the end node of the first and the start node of the second new link. Accordingly, all 
remaining nodes of the original link up to the new node become part of the first new link and all 
others part of the second new link.

\begin{figure}[!htb]
  \centering
  \includegraphics[width=1.0\textwidth]{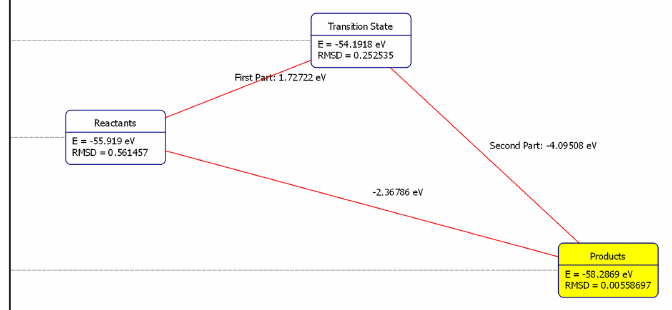}
  \caption{An exemplary reaction network graph in the Global Reactivity Monitoring App is shown. A 
possible transition state structure has been extracted by splitting the recorded link between the 
reactants and products nodes. The currently active node is highlighted by the program.
  \label{fig:grnetworkgraph}}
\end{figure}

\section{Conclusions}

The interactive study of chemical reactivity gives rise to new challanges unprecedented in quantum
chemistry due to the instantaneous interaction of the operator with the molecular system. To meet
them, we started with a careful analysis of the manipulations that the operator applies to the 
molecular system. The coupling of the haptic (force-feedback) device that is used for the 
manipulation is decisive for an interactive study of chemical reactivity as the force exerted on an 
atom of a reactant is the descriptor of reactivity in a reacting molecular system. A transparent and 
precise control over the atom position is of prime importance in reactivity studies and thus 
requires a direct coupling of the device to the manipulated atom(s). 

We classified the interactive exploration of chemical reactivity as local and global reactivity 
explorations. For a local reactivity exploration we identified the perpendicular forces and the 
evasive adaptation behavior as the main problems that may occur during an interactive exploration. 
The perpendicular forces may make it difficult for the operator to follow a specific path through 
the configuration space. The evasive adaptation behavior encountered when probing reactivity may 
hinder the user to reach certain configurations. For both effects, we proposed several schemes
that can help to achieve a desired behavior. They include the separate scaling of the perpendicular 
force component during an exploration and the adaptation of the rate of the continuously running 
structure optimization. 

In the last part of this work, we described an implementation of the ideas developed as Apps that 
can be loaded into the {\sc Samson} program environment. They allow the operator to interactively 
manipulate a molecular system with an haptic device, to study the system's reactivity, and to deal 
with difficult scenarios that he/she may encounter. 

Our DFTB App is an implementation of the standard non-self-consistent variant of DFTB that allows
other Apps in {\sc Samson} to calculate energies and forces for a given molecular structure. We could show 
that the DFTB method is an option for the reliable real-time computation of energies and forces for 
the interactive study of chemical reactivity. It is able to reliably predict molecular structures 
and to provide a satisfactory description of the reactivity for many molecular systems, while still 
being fast enough for providing energies and forces in real time. The DFTB method, in combination 
with the 3OB parameter set, is a useful tool for the haptic exploration of systems containing H, C, 
N, O, P and S atoms.

The Local Reactivity App together with the Phantom Direct App provide the functionality needed for 
local reactivity studies. The former enables the operator to control the continuously running 
structure optimization as well as the interaction with the molecular system with the haptic device. 
All the necessary parameters can be tuned through a readily accessible graphical user interface. 

The functionality of the Global Reactivity Monitoring App allows the operator to easily store, 
organize and analyze reaction networks. The nodes and links, corresponding to the stable 
intermediates and the interconnecting reaction paths, are visualized in a reaction network graph. 
The App allows one to keep track of an ongoing reaction network exploration. 

All these Apps extend the {\sc Samson} program framework to a virtual environment in which chemists are 
able to intuitively and interactively explore chemical reactivity of molecular systems. It provides 
all means necessary to interact with the molecular system, obtain the response upon the 
manipulations and explore the possible reaction paths. Our further development will aim at 
extending the range of quantum mechanical methods to calculate the energies and forces in real time. 

\section{Acknowledgment}

MPH and MR acknowledge support by ETH Zurich (grant number: ETH-08 11-2). MB and SR acknowledge  
funding from the European Research Council under the European Union's Seventh Framework Program 
(FP/2007-2013) / ERC Grant Agreement n. 307629

\appendix

\section{Technical Peculiarities}

As the feasibility of interactive reactivity exploration depends crucially on an efficient 
implementation, we shall provide these details in this appendix. 

\subsection{Local Reactivity App}

\begin{figure}[!htb]
  \centering
  \includegraphics[width=\textwidth]{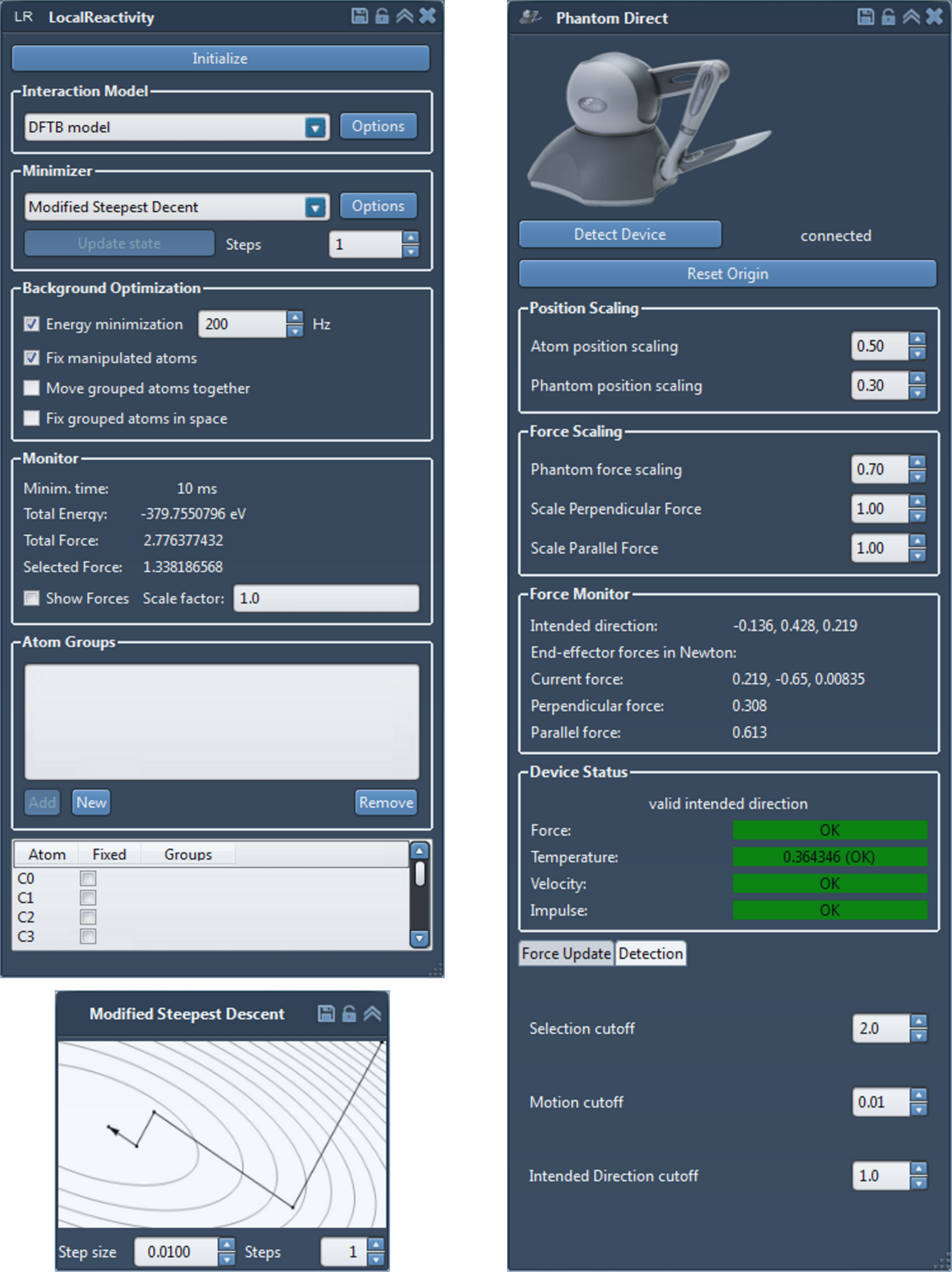}
  \caption{GUI elements for the two Apps devoted to the study of local reactivity. On the
left the Local Reactivity App window and the window for adjusting the steepest descent procedure
are shown. On the right the window of the Phantom Direct App is presented (the inlay depicts the 
haptic pointer device).
  \label{fig:localreactivitygui}}
\end{figure}

The main task of the Local Reactivity App is to provide access to all parameters necessary to tune 
the process of the background structure optimization. Additional visual elements such as the display 
of force vectors can also be switched on. Moreover, in the current version it serves also as an 
entry point for reactivity studies, as it takes care of all required initialization tasks such as 
the set up the parameters for the real-time response calculation, which is the electronic structure 
optimization and the force calculation. The functionality is implemented in the classes 
\texttt{SMA\-Local\-Reactivity} and \texttt{SMA\-Local\-Reactivity\-GUI} (see Table 
\ref{tbl:LRClasses}). The corresponding GUI elements are shown in Figure \ref{fig:localreactivitygui}. 

\begin{table}[!htb]
\centering
  \caption{Selection of the classes and their most important members that implemented the
functionality of the Local Reactivity App. }
\footnotesize
  \label{tbl:LRClasses}
  \begin{tabularx}{\textwidth}{llX} \toprule
    \multicolumn{3}{l}{\texttt{class SMALocalReactivity}}                                         \\
      & \texttt{updateState()}          &  Calls the structure optimizer which performs a number 
                                           of optimization cycles specified in the optimizer
                                           GUI panel.                                       \\
      & \texttt{show/hideForces()}      &  Switches the display of the forces on or off.          \\
      & \texttt{updateMonitor()}        &  Updates all the monitoring panels in the GUI.    \\
      & \texttt{addSimulator()}         &  Loads the structure optimizer chosen in the GUI. \\
      & \texttt{applyInteractionModel()}&  Loads the interaction model specified in the
                                           GUI.                                             \\
\cmidrule{1-3}                                                                              
    \multicolumn{2}{l}{\texttt{class SMALocalReactivityGUI}} &                                    \\
      & \texttt{simulationTimer}        & \texttt{QTimer} object that executes the function
                                          \texttt{optimize()} with a frequency given in the
                                          GUI (Local Reactivity State Update loop).         \\
      & \texttt{optimize()}             & Calls \texttt{SMA\-Local\-Reactivity::update\-State()}  \\
    \bottomrule
  \end{tabularx}
\end{table}

To start with an interactive exploration the operator needs to specify the particles of the 
molecular system under study. This is taken care of by the {\sc Samson} framework, which offers several 
convenient ways to set up a molecular structure. For instance, the operator can import a Cartesian
coordinate file by just pulling it on the main {\sc Samson} window. A click on ``Initialize'' internally 
connects the Local Reactivity App with the molecular structure and sets up the selected electronic 
structure method for calculating the energy and forces. In the current version the operator can 
choose between the ASED-MO method \cite{bosson2012} and the (non-self-consistent) DFTB 
implementation reported in this work. They are accessed over a common interface which allows the 
operator to exchange them by any other interaction model (implemented in a {\sc Samson} App) that can also
provide energies and forces in real time. 

With the continuously updated ``Monitor'' panel the operator can follow some overall properties of
the molecular system like the total energy, the magnitude of the total force, and the magnitude of 
the force on the currently selected atom. Also the forces on all atoms can be displayed as green 
arrows together with the molecular structure. To be able to detect small residual forces, the force 
vectors can be scaled. The ``Atom Groups'' panel allows the operator to define groups of atoms, 
which can behave as rigid bodies or can be fixed in space. The table at the bottom of the App 
window provides an overview over all atoms and indicates whether they are fixed or part of a group. 
It is also possible to directly fix the position of single atoms in this table. These features may 
help to circumvent the problem of evasive adaptations.

In the ``Background Optimization'' panel the operator controls the parameters of the structural
response calculation. It is possible to start and stop the background energy minimization and to
adapt the frequency of its optimization loop. The operator can exclude the manipulated atom from the
optimization by fixating it to the position given by the input device. Also, atoms united in a group
can be moved together by applying the same displacement to all atoms in the group. Note, however,
that this does not necessarily correspond to rigid body movement, since the atoms in the group may
relax independently. Also these features may help the operator to avoid unwanted evasive adaptation 
of the manipulated molecules. 

The loop for the background optimization of the molecular structure is implemented in
\texttt{SMALocal\-Reactivity::update\-State()}, which is indirectly called by the
\texttt{simulation\-Timer} object in the \texttt{SMA\-Local\-Reactivity\-GUI} class (see
Table \ref{tbl:LRClasses}). The optimizer frequency specified in the GUI is taken to set up
the \texttt{simulation\-Timer}. Since the background optimization loop is part of the GUI
thread (see Figure \ref{fig:threadsandapps}), this optimization is not guaranteed to run at the
specified frequency. Thus, a frequency below $100\,\mathrm{Hz}$ is unlikely to be realized
considering that this already corresponds to an execution every $10\,\mathrm{ms}$ and recalling that
the GUI thread is also responsible for processing the events created by all other GUI
elements.

\subsection{Phantom Direct App}

\begin{figure}[!htb]
  \centering
  \includegraphics[width=\textwidth]{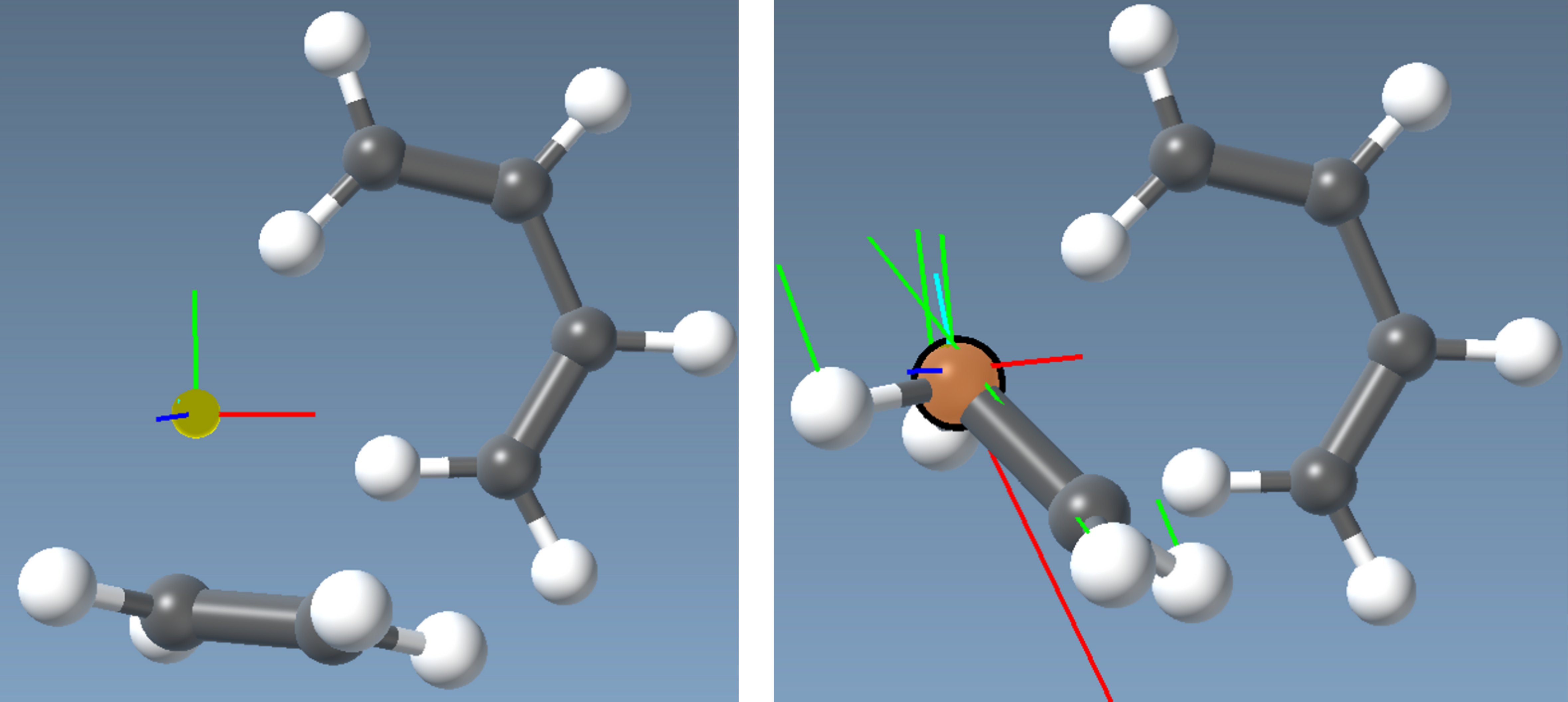}
  \caption{On the left, the molecular structure visualization in {\sc Samson} and the graphical 
representation of the HIP (as a yellow ball with three axes) is shown before the operator 
interacts with the system. On the right, the operator is manipulating an atom (highlighted in 
orange), which causes the other atoms to rearrange by following their forces (in green). In red, 
the force experienced by the operator and in turquoise the intended direction of manipulation is 
depicted.
  \label{fig:moleculesAndHIP}}
\end{figure}

With a click on the ``Detect Device'' button the device loop is started and a representation of the
end-effector position and orientation, the haptic interpoint point (HIP), appears (see Figure 
\ref{fig:moleculesAndHIP}). To interact with the molecular system the operator has to position the 
HIP close to an atom and press the device button. If the HIP is close to an atom while the button is 
pressed the operator can manipulate the atom and simultaneously experiences the force on this atom. 
Such a situation is depicted in Figure \ref{fig:moleculesAndHIP} where on the left hand side the HIP 
representation is shown and on the right hand side the operator is manipulating the position of one 
atom of the ethene molecule. During the manipulation the ``Force Monitor'' panel shows the intended 
direction, the currently rendered force and the force decomposition in the parallel and 
perpendicular (to the intended direction) component. Additionally, the intended direction is 
displayed as a turquoise arrow originating from the HIP. If a force is exerted on the manipulated 
atom, this force is displayed as a red arrow with its origin at the HIP. The overall scaling of the 
force magnitude and the scaling of the components can be adjusted in the ``Force Scaling'' panel. 
The ``Position Scaling'' panel allows adjusting the scale factors that affect the transformation of 
the end-effector position to the position of the HIP in the VE. In addition, the operator can adjust 
several cut-off values for the detection of movements and the atom selection. Also the damping 
factor for the force update can be modified there.

The cascade of transformations is
implemented in the class \texttt{SMA\-Phantom\-Connector\-Direct}. The parameters for the
transformations are taken from the Phantom Direct GUI (see Figure \ref{fig:localreactivitygui}
and class \texttt{SMA\-Phantom\-Connector\-Direct\-GUI} in Table \ref{tbl:PDClasses}).

\begin{table}[!htb]
\centering
  \caption{Selection of the classes and their most important members that implement the
functionality of the Phantom Direct App. The haptic interaction point (HIP) is the position of the 
haptic device (HD) end-effector in the virtual environment.}
\footnotesize
  \label{tbl:PDClasses}
  \begin{tabularx}{\textwidth}{llX} \toprule
    \multicolumn{3}{l}{\texttt{class SMAPhantomConnectorDirect}}                              \\ 
      & \texttt{phantomCurrentTransform} & Object that stores position and orientation of the
end-effector in the world frame.                                                              \\
      & \texttt{detectDevice()}         & Set up device and start scheduler.                  \\
      & \texttt{detectClosestAtom()}    & Detect if an atom is close to the HIP.        \\
      & \texttt{getPhantomTransformInWorldFrame()} & Returns position and orientation of the HIP
       in the molecular (world) frame.                                                     \\
      & \texttt{moveSelectedAtom()}     & Move active atom to current HIP position.    \\
      & \texttt{resetOrigin()}          & Reset origin for the transformation.                \\
      & \texttt{updateState()}          & Callback function of HD loop.                 \\ 
\cmidrule{1-3}  
    \multicolumn{3}{l}{\texttt{class SMAPhantomConnectorDirectGUI}}                           \\
      & \texttt{simulationTimer}        & \texttt{QTimer} object that executes every
                                          $10\,\mathrm{ms}$ the function
                                          \texttt{move\-Selected\-Atom()}.                    \\
      & \texttt{moveSelectedAtom()}     & Calls \texttt{move\-Selected\-Atom()} in
                                          \texttt{SMA\-Phantom\-Connector\-Direct}.           \\
\cmidrule{1-3} 
    \multicolumn{3}{l}{\texttt{class SMAPhantomConnectorDirectHIP}}                           \\
      & \texttt{display()}              & Executes OpenGL commands to render a sphere and a
coordinate system representing position and orientation of the HIP.                     \\
    \bottomrule
  \end{tabularx}
\end{table}

The GUI also provides extensive monitoring of the force rendered by the device (in contrast to the
molecular force monitored in the Local Reactivity App) and of the haptic device status. The App is
specifically designed for the Phantom Desktop device from Sensable Inc \cite{sensable}. It
communicates with the device via an application programming interface (API) that is part of the 
OpenHaptics$^\text{\textregistered}$ toolkit provided by Sensable \cite{openhapticstoolkit}. This 
haptic device API (HDAPI) provides a scheduler that starts a loop in a separate high-priority thread 
for the communication with the haptic device. This device loop runs at a frequency of $1\,\mathrm{kHz}$ 
and is responsible for rendering the forces and obtaining the position and orientation of the 
end-effector (the point corresponding to the tip of the pen-like device). The tasks that read or 
write the device state are implemented as callback functions registered in the scheduler. 

The main functionality of the Phantom Direct App is executed in two simultaneously running loops
that are part of two different threads (see Figure \ref{fig:threadsandapps}). The functionality of the
App is implemented in the two \texttt{SMA\-Phantom\-Connector\-Direct} classes
(Table \ref{tbl:PDClasses}). The device loop is implemented in the \texttt{update\-State()} function
called by the device scheduler, whereas the simulation loop is implemented in
\texttt{move\-Selected\-Atom()} called by the \texttt{simulationTimer}. Upon clicking on the
``Detect Device'' button the \texttt{detect\-Device()} function initializes the haptic device and
starts the scheduler that handles the device loop. Subsequently the position and orientation of the
end-effector is stored as the reference position. This can be also issued by clicking on ``Reset
Origin'' which effectively resets the origin of the haptic device. With scheduling the callback
function \texttt{update\-State()} the device loop is started. The final steps of the initialization
process are to create the visual representation of the HIP in the virtual environment, to enable all 
user interface elements and to start the simulation loop by calling \texttt{move\-Selected\-Atom()} 
for the first time.

\begin{figure}[!htb]
  \centering
  \includegraphics[width=0.8\textwidth]{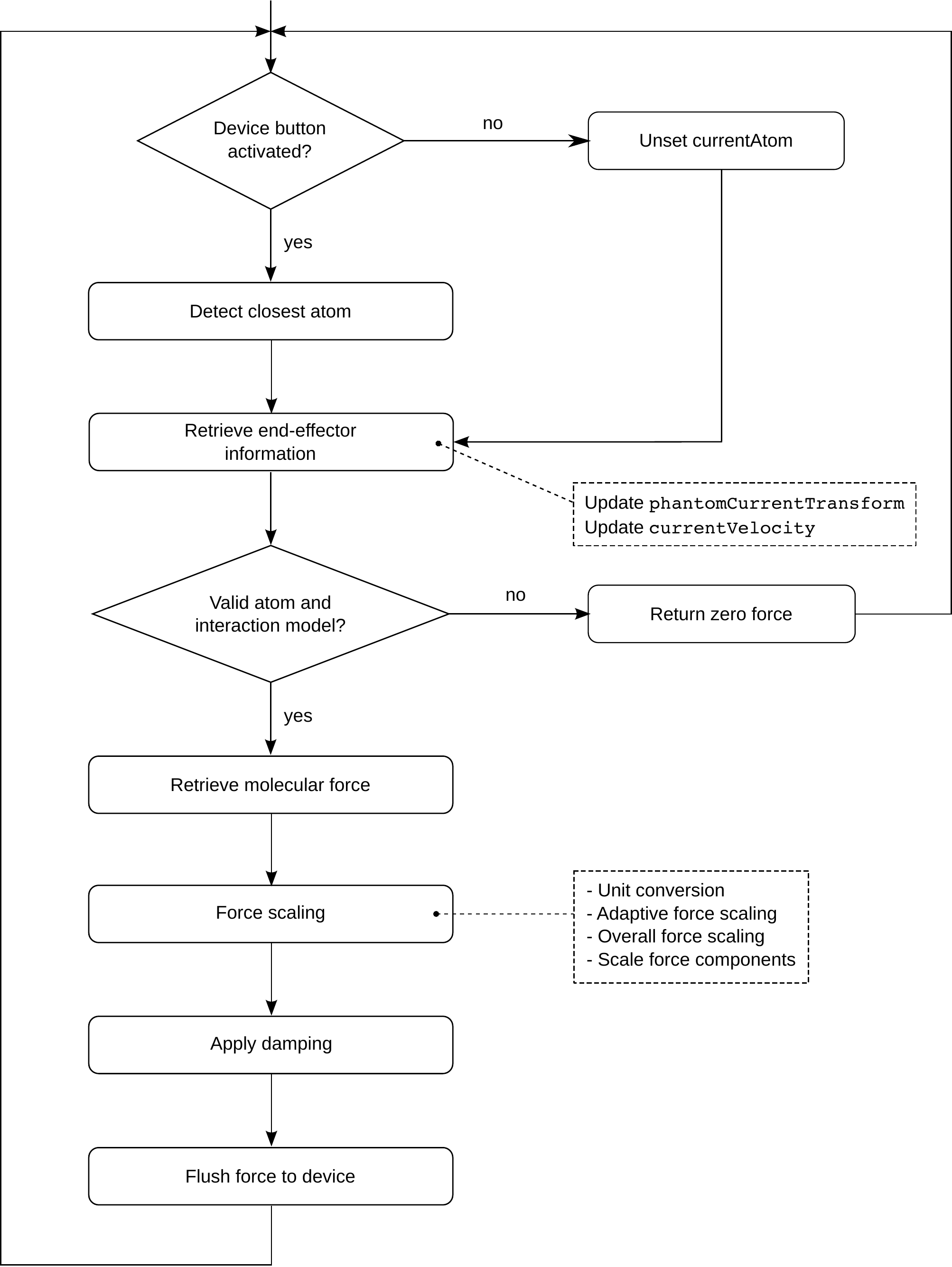}
  \caption{Flow chart of the function \texttt{SMA\-Phantom\-Connector\-Direct::update\-State()}
  called at every pass of the device loop maintained by the haptic device scheduler.
  \label{fig:flowchartDeviceLoop}}
\end{figure}

The device loop reads the current position and orientation (\texttt{phantom\-Current\-Transform}) of
the device end-effector and sends a force to the device when the operator is manipulating an atom.
Unless otherwise stated, the functions and objects accessed from within the loop are also members of
the class \texttt{SMA\-Phantom\-Connector\-Direct}. Figure \ref{fig:flowchartDeviceLoop} provides an
overview over the procedure executed in \texttt{update\-State()} in a flow-chart.

In the beginning the function determines by reading the state of the device button whether the
operator tries to manipulate an atom or whether the device is just moved. If an atom atom is to be
moved, \texttt{detect\-Closest\-Atom()} will search for an atom within a certain radius around the
HIP position. In case of success, a reference to the atom is stored in \texttt{currentAtom}.
The next step is to obtain position, orientation and velocity of the end-effector of the haptic
device. The velocity is stored in \texttt{current\-Velocity} and the
\texttt{phantom\-Current\-Transform} object is updated with the positional and orientational
information. 

The function \texttt{get\-Phantom\-Transform\-In\-World\-Frame()} allows us accessing the current
position and orientation of the HIP in the reference frame of the VE. In this
transformation step the atom position scaling factor is applied. A small factor effectively slows
down the movements performed by the operator. As discussed in the preceding sections, this can be
exploited to allow for adiabatic manipulations in those cases where the background optimization is 
not fast enough. A small factor may also help to reduce oscillatory movements due to fast changes 
in the force direction as they can occur close to steep minima. Only when a force is provided 
(requires a valid current atom and a valid interaction model) does the function continue with the 
force processing, otherwise it returns to the scheduler without rendering a force. 

The first step in the sequence of transformations is to convert the molecular force into SI units. 
The forces provided by the interaction model in {\sc Samson} are in electron volts per \r{A}ngstr\"om. 
Since the conversion factor to SI unit  
\begin{align}
    1\,\mathrm{eV}\,\text{\AA}^{-1} = 1.60218\times 10^{-9}\,\mathrm{N}
\end{align}
is just a constant, it trivially obeys the rules stated above. In the so-called adaptive
force scaling step the magnitude of the force vector is adjusted to be in the range which can be
rendered by the device. To exploit the given range in an optimal way, small forces are scaled up
whereas very high forces are scaled down \cite{bolopion2011}. This allows the operator to feel the
subtle details of an almost flat surface while simultaneously very high forces do not force the
operator to perform unwanted movements. An arc tangent function of the form
\begin{align}\label{eq:adaptivescaling}
  f_\mathrm{out}(f_\mathrm{in}) = \frac{2\,f_\mathrm{max}}{\pi} \arctan(m\,f_\mathrm{in}) 
\end{align}
is chosen to scale the magnitude of the force. It is similar to the one chosen in
Ref.~\cite{bolopion2011} but ours is of a simpler form involving less parameters.
Here, $f_\mathrm{out}$
is the force magnitude after the adaptive scaling and $f_\mathrm{in}$ is the input force
magnitude after the unit conversion. The parameter $f_\mathrm{max}$ is the force limit of the
device and $m$ determines the ``slope'' of the function. By adjusting $m$ and $f_\mathrm{max}$, one
can change the turning point from magnification to reduction (i.e., where the first derivative is
$1.0$) of the input force. In Figure \ref{fig:scalingsDampingsPlot} (bottom left plot), the graph of 
this function is shown for suitable parameters applied in our implementation. The strictly 
monotonically increasing and strictly positive function fulfills our requirements for a conversion 
function and thus assures that the transformation does not introduce any artificial features into 
the PES. By only scaling the magnitude of the force, it also does not change the direction of the 
force vector. 

The haptic interaction point is implemented in class \texttt{SMA\-Phantom\-Connector\-Direct\-HIP} 
in the \texttt{display()} function. This function is continuously called by the {\sc Samson} rendering 
engine and executes some simple OpenGL commands to render a sphere and three perpendicular axes 
representing the position and orientation of the HIP (see Figure \ref{fig:moleculesAndHIP}). 
Position and orientation of the HIP in the correct reference frame are obtained by calling the 
function \texttt{get\-Phantom\-Transform\-In\-World\-Frame()}. 

The actual displacement of an atom, when manipulated by the haptic device, is done in the separate 
simulation loop (see Figure \ref{fig:threadsandapps}). This separation is necessary, since the device 
loop runs at a too high frequency. The simulation loop checks whether a manipulation is currently 
performed and which atom is affected. It reads the position of the HIP by calling the function 
\texttt{get\-Phantom\-Transform\-In\-World\-Frame()}. This position is then taken to determine the 
new position for the atom. In addition to move the selected atom, the simulation loop is responsible 
for monitoring by updating all GUI monitoring elements.

The functionality of the simulation loop is implemented in the function
\texttt{move\-Selected\-Atom()} of class \texttt{SMA\-Phantom\-Connector\-Direct} and is triggered
by the \texttt{simulation\-Timer} object (see Table \ref{tbl:PDClasses}). Upon execution, the function
performs the following procedure: It starts by checking whether an atom is currently manipulated
and, if this is the case, the current position of the atom and the haptic interaction point in the
world frame is retrieved. From this, the displacement can be calculated and checked against the
specified motion threshold. If the displacement is above the threshold, it is scaled by the atom
position scaling factor specified in the GUI and eventually applied to the atom. After that,
all user interface elements are updated. 

\subsection{Global Reactivity Monitoring App}

\begin{table}[!htb]
\footnotesize
\centering
  \caption{Selection of the classes and their most important members that are part of the Global
Reactivity Monitoring App.}\label{tbl:GRClasses1}
  \begin{tabularx}{\textwidth}{llX} \toprule
    \multicolumn{3}{l}{\texttt{class SMAGlobalReactivityGUI}}                              \\
      & \texttt{showNetworkView()}       & Starts the network view. \\
\cmidrule{1-3}  
    \multicolumn{3}{l}{\texttt{class GRNetworkView}}                              \\
      & \texttt{\_timer}                   & \texttt{QTimer} object that executes every
                                          $100\,\mathrm{ms}$ the function
                                          \texttt{update\-Network\-Elements()}.  \\
      & \texttt{\_model}                  & \texttt{GRNetworkDataModel} objects that stores the
                                           data model.\\
      & \texttt{\_graph<View/Scene>}      & \texttt{QGraphicsScene} and \texttt{QGraphicsView} pair
                                           for the graph display.\\
      & \texttt{updateNetworkElements()} & Forces all elements of the data model to update
                                           themselves. \\
      & \texttt{onStructuralSignal()}    & Reacts to structural changes in case of an active path
                                           recording. \\
      & \texttt{<open/save/$\ldots$>Session()}     & Functions responsible for session management. \\     
      & \texttt{add<Node/Link>()}        & Adding a node or link item to the data model stored in 
                                           \texttt{\_model}\\    
    \bottomrule
  \end{tabularx}
\end{table}

\begin{figure}[!htb]
  \centering
  \includegraphics[width=1.0\textwidth]{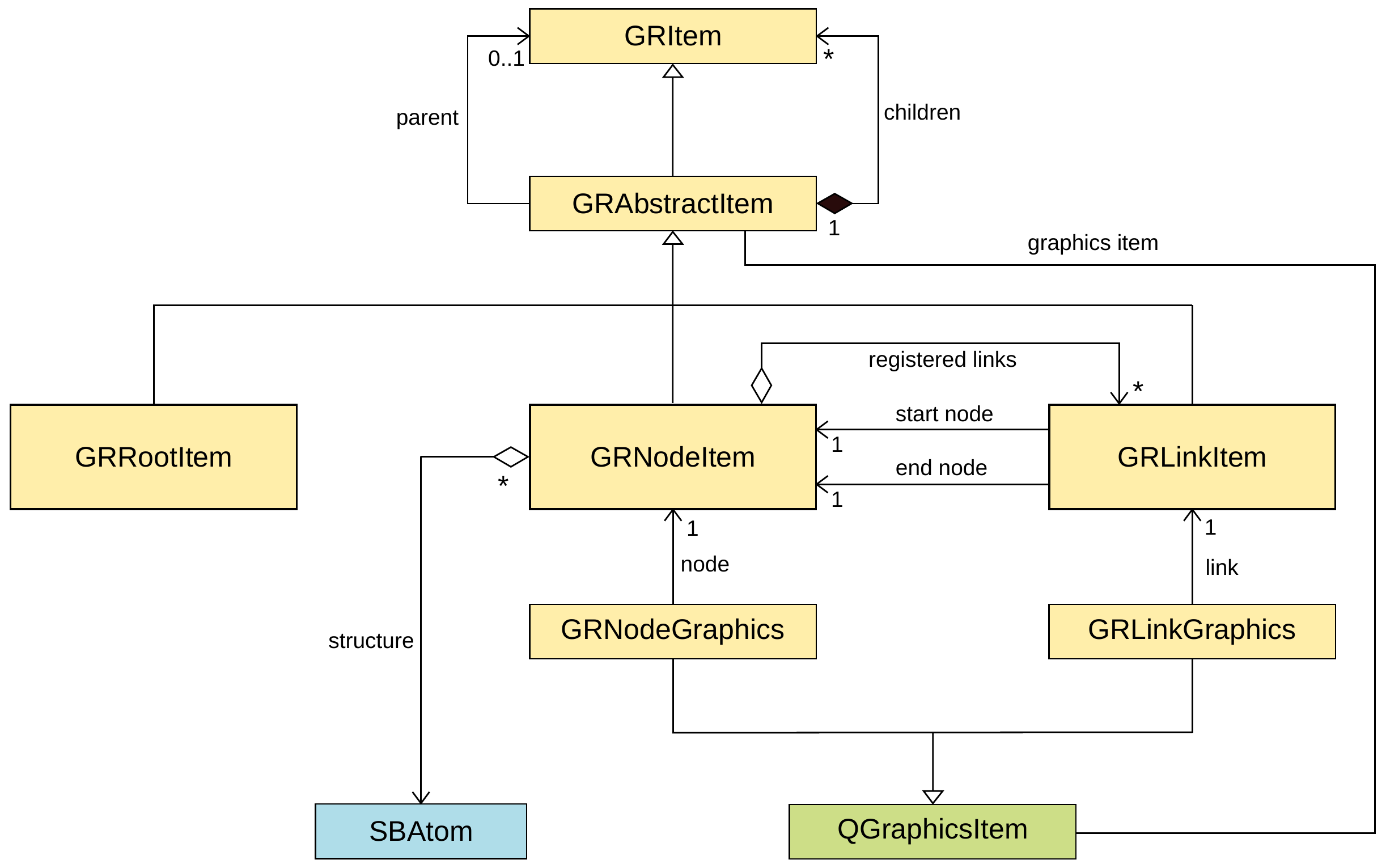}
  \caption{Class diagram for the main components in the global reactivity data model employing the
notation defined in the Unified Modeling Language (UML). The classes that are part of the global 
reactivity data model are yellow, classes from {\sc Samson} are blue and Qt classes are green. This 
diagram does not show that \texttt{GRLink\-Graphics} in fact derives from 
\texttt{QGraphics\-Path\-Item} which is in turn derived from \texttt{QGraphics\-Item}. 
\texttt{GR\-Abstract\-Item} provides default implementations of the \texttt{GRItem} interface.
  \label{fig:grdatamodel}}
\end{figure}

The structure of reaction network maps suggests a data model resembling the 
composite design pattern \cite[163--174]{gamma1995comppatt} (see Figure \ref{fig:grdatamodel} and
Table \ref{tbl:GRClasses2}). The classes storing the data of nodes and the links are called
\texttt{GRNode} and \texttt{GRLink}, respectively. \texttt{GRItem} is the common interface from
which both classes are derived. It provides all function signatures needed to access the child items
of a composite class and to set (and get) the name and the associated graphics object (either
\texttt{GRNode\-Graphics} or \texttt{GRLink\-Graphics}). The graphics objects are responsible for
the graphical representation of an item in the graph view. The graph view itself is implemented
employing Qt's \texttt{QGraphics\-View} and \texttt{QGraphics\-Scene} classes which is why the
classes \texttt{GRNode\-Graphics} and \texttt{GRLink\-Graphics} are both derived from
\texttt{QGraphicsItem}. 

\begin{table}[!htb]
\footnotesize
\centering
  \caption{Selection of the classes and their most important members that constitute the data model
of the Global Reactivity Monitoring App. See also Figure \ref{fig:grdatamodel} for an overview over
all classes.}
  \label{tbl:GRClasses2}
  \begin{tabularx}{\textwidth}{llX} \toprule
    \multicolumn{3}{l}{\texttt{class GRNetworkDataModel}}                                 \\
      & \texttt{rootItem}                & The invisible root item that has all top-level items as 
                                           children.\\
      & \texttt{append<Node/Link>()}     & Add a top-level node or link item to the data model,
                                           i.e., make it a children of \texttt{rootItem}. \\
      & \texttt{remove<Node/Link>()}     & Remove a top-level node or link item from the data model,
                                           i.e., remove it from \texttt{rootItem}. \\
      & \texttt{load()/save()}           & Loading and saving of the data items from and to
                                           XML files.\\
\cmidrule{1-3} 
    \multicolumn{3}{l}{\texttt{class GRNode}} \\
      & \texttt{links()}                 & Get references to all links point to or originating from
                                           the node. \\
      & \texttt{<add/remove>Link()}      & Register and deregister links. \\
      & \texttt{setAtoms()/atoms()}      & Access stored structure snapshot.\\
      & \texttt{setRmsd()/rmsd()}        & Access RMSD value. \\
      & \texttt{setEnergy()/energy()}    & Access stored energy. \\
      & \texttt{setActive()/active()}    & Access activity flag. \\
      & \texttt{checkStructure()}        & Determine RMSD value.\\
      & \texttt{setVector()/vector()}    & Access vector in configuration space.\\
\cmidrule{1-3} 
    \multicolumn{3}{l}{\texttt{class GRLink}} \\
      & \texttt{deltaE()}                & Access energy difference between start and end node. \\
      & \texttt{setFromNode()/fromNode()}& Access start node. \\
      & \texttt{setToNode()/toNode()}   & Access end node. \\
      & \texttt{update()}                & Determine energy difference and update graphics item.\\
    \bottomrule
  \end{tabularx}
\end{table}

When the network window is opened the operator starts with an empty graph. The program has, however,
already created an invisible (without a graphics object) root item to store the data objects. Upon
addition of a new item to the data model the corresponding graphics item is added to the scene. The
graphics item is informed about state changes by its data item to keep the graphical
representation synchronized with the underlying data model. All items that are visible are stored as 
direct children of the invisible root item (also called top-level items). The reaction network 
hierarchy implies the following rules, that are enforced by the implementation:
\begin{itemize}
 \item \texttt{GRItem}s cannot be child of more than one \texttt{GRItem} (including the invisible
       root item). 
 \item \texttt{GRLink}s can only be top-level items.
 \item \texttt{GRNode}s can be top-level items (with a graphics item) or part of other top-level
        items (without a graphics item).
 \item \texttt{GRNode}s cannot be parents of \texttt{GRLink}s, but of other \texttt{GRNode}s.
\end{itemize}
Although in principle nodes and links can have \texttt{GRNode}s as children, this feature is
currently only implemented for the \texttt{GRLink}s. The children of \texttt{GRLink} items are nodes
representing the elements of the path. In contrast to top-level items the \texttt{GRNode}s that are
path elements do not have a graphics object. 

The difference between \texttt{GRNode}s and \texttt{GRLink}s manifests in their fields and methods
additional to the ones inherited from \texttt{GRItem}. Figure \ref{fig:grdatamodel} shows that
\texttt{GRLink}s and \texttt{GRNode}s are in fact not directly derived from \texttt{GRItem} but from
an intermediate abstract class (\texttt{GRAbstract\-Item}) that provides the default implementation
of the functions defined in \texttt{GRItem}.

Every \texttt{GRNode} stores a snapshot of the molecular structure as a collection of
\texttt{SBAtom}s (a {\sc Samson} data type representing atoms). In addition, the nodes update the
RMSD between the stored structure and the current structure in {\sc Samson}. The continuous
calculation of the RMSD is again implemented employing a \texttt{QTimer} that forces the items
to update their state every $100\,\mathrm{ms}$. Before calculating the RMSD of a node the
structure is fitted to the structure in {\sc Samson} employing a quaternion fit method \cite{kneller1991}.
This ensures that structures, which differ only in overall orientation and translation, are detected
as equal. Based on the RMSD an additional flag is continuously updated which indicates that 
the structure of the node is equal (within a certain threshold) to the current structure in {\sc Samson}.
Additionally, the energy, a vector representation of the structure, and references to the links
connected with the node are stored. 

The \texttt{GRLink}s store less information than the \texttt{GRNode}s, since their main information
content is the path constituted by their children. They keep references to the nodes where they
start and end and save the energy difference between the two nodes. 

A node can be created from every structure that is currently active in {\sc Samson}. The current
structure and energy is saved in a new \texttt{GRNode} object, which is then made child of the
root item. In addition, a \texttt{GRNodeGraphics} object is created, assigned to the \texttt{GRNode}
object and registered, their name can be changed by a double click, and a double click with the
middle mouse button sets the current structure in {\sc Samson} to the structure stored in the node. With
this the operator can quickly navigate through the network of nodes and organize it. 

The results of a global reactivity session can be stored in the Extensible Markup Language (XML) 
format. All nodes and links will be stored and can be loaded to continue an already started 
exploration session. To exchange path elements between different sessions, also single nodes or 
links can be exported to an XML file and also imported from it. In addition, nodes can also be saved 
in the xyz file format. This eases the export and import of the structures found from and into other 
quantum chemistry programs, where they can be refined if deemed necessary.


\begin{thebibliography}{10}

\bibitem{dykstra2005}
C.~Dykstra,\ G.~Frenking,\ K.~Kim,\ G.~Scuseria,\ (Eds.),  \textit{Theory and
  Applications of Computational Chemistry: The First Forty Years;} Elsevier:
  2005.

\bibitem{goedecker1999}
S.~Goedecker, \textit{Rev. Mod. Phys.} \textbf{1999,} \textsl{71,} 1085-1123.

\bibitem{rubensson2011}
E.~H. Rubensson,\ E.~Rudberg,\ P.~Salek,  Methods for Hartree--Fock and Density
  Functional Theory Electronic Structure Calculations with Linearly Scaling
  Processor Time and Memory Usage.   In  \textit{Linear-Scaling Techniques in
  Computational Chemistry and Physics}, Vol.~13; R.~Zalesny,\ M.~G.
  Papadopoulos,\ P.~G. Mezey,\ J.~Leszczynski,\ (Eds.),  Springer Netherlands:
  2011.

\bibitem{ochsenfeld2007}
C.~Ochsenfeld,\ J.~Kussmann,\ D.~S. Lambrecht,  Linear-Scaling Methods in
  Quantum Chemistry.   In  \textit{Reviews in Computational Chemistry}; John
  Wiley \& Sons, Inc.: 2007.

\bibitem{bosson2011}
M.~Bosson,\ S.~Grudinin,\ X.~Bouju,\ S.~Redon, \textit{J. Comput. Phys.}
  \textbf{2011,} \textsl{231,} 2581-2598.

\bibitem{hanwell2012}
M.~D. Hanwell,\ D.~E. Curtis,\ D.~C. Lonie,\ T.~Vandermeersch,\ E.~Zurek,\
  G.~R. Hutchison, \textit{J. Cheminform.} \textbf{2012,} \textsl{4,} 1-17.

\bibitem{marti2009}
K.~H. Marti,\ M.~Reiher, \textit{J. Comput. Chem.} \textbf{2009,} \textsl{30,}
  2010--2020.

\bibitem{haag2011}
M.~P. Haag,\ K.~H. Marti,\ M.~Reiher, \textit{ChemPhysChem} \textbf{2011,}
  \textsl{12,} 3204-3213.

\bibitem{bosson2012}
M.~Bosson,\ C.~Richard,\ A.~Plet,\ S.~Grudinin,\ S.~Redon, \textit{J. Comput.
  Chem.} \textbf{2012,} \textsl{33,} 779-790.

\bibitem{bosson2013}
M.~Bosson,\ S.~Grudinin,\ S.~Redon, \textit{J. Comput. Chem.} \textbf{2013,}
  \textsl{34,} 492--504.

\bibitem{haag2013}
M.~P. Haag,\ M.~Reiher, \textit{Int. J. Quantum Chem.} \textbf{2013,}
  \textsl{113,} 8-20.

\bibitem{haag2014a}
M.~P. Haag,\ M.~Reiher, \textit{Faraday Discuss.} \textbf{2014,} \textsl{169,
  DOI: 10.1039/C4FD00021H,} DOI: 10.1039/C4FD00021H.

\bibitem{brooks1990}
F.~P. Brooks,~Jr.,\ M.~Ouh-Young,\ J.~J. Batter,\ P.~Jerome~Kilpatrick,
  \textit{SIGGRAPH Comput. Graph.} \textbf{1990,} \textsl{24,} 177-185.

\bibitem{krenek1999}
A.~K\v{r}enek,\ M.~\v{C}ernohorsk\'{y},\ Z.~Kabel\'{a}c,\ Z.~Kabel\'{a}\v{c},
  Haptic Visualization of Molecular Model.   In  \textit{WSCG'99 - The 7-th
  International Conference in Central Europe on Computer Graphics,
  Visualization and Interactive Digital Media'97}, Vol.~1-3; V.~Skala,\ (Ed.),
  Univ.of West Bohemia Press: 1999.

\bibitem{harvey2000}
E.~Harvey,\ C.~Gingold,  Haptic representation of the atom.   In
  \textit{Information Visualization, 2000. Proceedings. IEEE International
  Conference on}; 2000.

\bibitem{comai2009}
S.~Comai,\ D.~Mazza,  A Haptic-Enhanced System for Molecular Sensing.   In
  \textit{Human-Computer Interaction --- INTERACT 2009}, Vol.~5727; T.~Gross,\
  J.~Gulliksen,\ P.~KotzÃ©,\ L.~Oestreicher,\ P.~Palanque,\ R.~Prates,\
  M.~Winckler,\ (Eds.),  Springer Berlin Heidelberg: 2009.

\bibitem{davies2009}
R.~A. Davies,\ J.~S. Maskery,\ N.~W. John,  Chemical Education Using Feelable
  Molecules.   In  \textit{Proceedings of the 14th International Conference on
  3D Web Technology}; Web3D '09 ACM: New York, NY, USA, 2009.

\bibitem{sensable}
{SensAble Technologies, Inc}, ``{Phantom Desktop Device}'',
  http://www.sensable.com, accessed on 2014/04/14,.

\bibitem{bolopion2010a}
A.~Bolopion,\ B.~Cagneau,\ S.~Redon,\ S.~R\'{e}gnier, \textit{J. Mol. Graph.
  Model.} \textbf{2010,} \textsl{29,} 280-289.

\bibitem{eyring1935}
H.~Eyring, \textit{J. Chem. Phys.} \textbf{1935,} \textsl{3,} 107--115.

\bibitem{olsson2006}
M.~H. Olsson,\ J.~Mavri,\ A.~Warshel, \textit{Philos. Trans. R. Soc., B}
  \textbf{2006,} \textsl{361,} 1417-1432.

\bibitem{qt}
{Digia plc}, ``Qt Project Reference Pages'',
  http://qt-project.org/doc/qt-5/reference-overview.html, accessed on
  2004/04/14,.

\bibitem{porezag1995}
D.~Porezag,\ T.~Frauenheim,\ T.~K\"ohler,\ G.~Seifert,\ R.~Kaschner,
  \textit{Phys. Rev. B} \textbf{1995,} \textsl{51,} 12947--12957.

\bibitem{seifert1996}
G.~Seifert,\ D.~Porezag,\ T.~Frauenheim, \textit{Int. J. Quantum Chem.}
  \textbf{1996,} \textsl{58,} 185--192.

\bibitem{elstner2014}
M.~Elstner,\ G.~Seifert, \textit{Philos. Trans. R. Soc., A} \textbf{2014,}
  \textsl{372,}.

\bibitem{elstner1998}
M.~Elstner,\ D.~Porezag,\ G.~Jungnickel,\ J.~Elsner,\ M.~Haugk,\
  T.~Frauenheim,\ S.~Suhai,\ G.~Seifert, \textit{Phys. Rev. B} \textbf{1998,}
  \textsl{58,} 7260--7268.

\bibitem{gaus2013}
M.~Gaus,\ A.~Goez,\ M.~Elstner, \textit{J. Chem. Theory Comput.} \textbf{2013,}
  \textsl{9,} 338-354.

\bibitem{gaus2014b}
M.~Gaus,\ X.~Lu,\ M.~Elstner,\ Q.~Cui, \textit{J. Chem. Theory Comput.}
  \textbf{2014,} \textsl{10,} 1518-1537.

\bibitem{slater1954}
J.~C. Slater,\ G.~F. Koster, \textit{Phys. Rev.} \textbf{1954,} \textsl{94,}
  1498--1524.

\bibitem{aradi2007}
B.~Aradi,\ B.~Hourahine,\ T.~Frauenheim, \textit{J. Phys. Chem. A}
  \textbf{2007,} \textsl{111,} 5678-5684.

\bibitem{numericalRecipes}
W.~Press,\ B.~Flannery,\ S.~Teukolsky,\ W.~Vetterling, \textit{Numerical
  Recipes in C: The Art of Scientific Computing;} Cambridge University Press:
  1992.

\bibitem{mkl}
``Math Kernel Library'',  See
  http://software.intel.com/en-us/articles/intel-mkl.

\bibitem{mayer1983}
I.~Mayer, \textit{Chem. Phys. Lett.} \textbf{1983,} \textsl{97,} 270 - 274.

\bibitem{gaus2011}
M.~Gaus,\ Q.~Cui,\ M.~Elstner, \textit{J. Chem. Theory Comput.} \textbf{2011,}
  \textsl{7,} 931-948.

\bibitem{becke1993}
A.~D. Becke, \textit{J. Chem. Phys.} \textbf{1993,} \textsl{98,} 5648-5652.

\bibitem{lee1988}
C.~Lee,\ W.~Yang,\ R.~G. Parr, \textit{Phys. Rev. B} \textbf{1988,}
  \textsl{37,} 785--789.

\bibitem{stephens1994}
P.~J. Stephens,\ F.~J. Devlin,\ C.~F. Chabalowski,\ M.~J. Frisch, \textit{J.
  Phys. Chem.} \textbf{1994,} \textsl{98,} 11623-11627.

\bibitem{dunning1989}
J.~T.~H.~Dunning, \textit{J. Chem. Phys.} \textbf{1989,} \textsl{90,}
  1007-1023.

\bibitem{weigend2005}
F.~Weigend,\ R.~Ahlrichs, \textit{Phys. Chem. Chem. Phys.} \textbf{2005,}
  \textsl{7,} 3297-3305.

\bibitem{raghavachari1989}
K.~Raghavachari,\ G.~W. Trucks,\ J.~A. Pople,\ M.~Head-Gordon, \textit{Chem.
  Phys. Lett.} \textbf{1989,} \textsl{157,} 479 - 483.

\bibitem{anderson1994}
A.~B. Anderson, \textit{Int. J. Quantum Chem.} \textbf{1994,} \textsl{49,}
  581-589.

\bibitem{diaz2001}
C.~Diaz,\ F.~Mendizabal, \textit{Bol. Soc. Chil. Qu\'{i}m.} \textbf{2001,}
  \textsl{46,} 293 - 299.

\bibitem{hehre1969}
W.~J. Hehre,\ R.~F. Stewart,\ J.~A. Pople, \textit{J. Chem. Phys.}
  \textbf{1969,} \textsl{51,} 2657-2664.

\bibitem{jacob2012}
C.~R. Jacob,\ M.~Reiher, \textit{International Journal of Quantum Chemistry}
  \textbf{2012,} \textsl{112,} 3661--3684.

\bibitem{ahlrichs1989}
R.~Ahlrichs,\ M.~B\"ar,\ M.~H\"aser,\ H.~Horn,\ C.~K\"olmel, \textit{Chem.
  Phys. Lett.} \textbf{1989,} \textsl{162,} 165-169.

\bibitem{sakai2000}
S.~Sakai, \textit{J. Phys. Chem. A} \textbf{2000,} \textsl{104,} 922-927.

\bibitem{gaussian09}
M.~J. Frisch, \textit{et al.}\  ``Gaussian~09'',  Gaussian Inc. Wallingford CT
  2009.

\bibitem{bandaranayake1980}
W.~M. Bandaranayake,\ J.~E. Banfield,\ D.~S.~C. Black, \textit{J. Chem. Soc.{,}
  Chem. Commun.} \textbf{1980,} \textsl{19,} 902-903.

\bibitem{rossi2007}
R.~Rossi,\ M.~Isorce,\ S.~Morin,\ J.~Flocard,\ K.~Arumugam,\ S.~Crouzy,\
  M.~Vivaudou,\ S.~Redon, \textit{Bioinformatics} \textbf{2007,} \textsl{23,}
  i408--i417.

\bibitem{openhapticstoolkit}
{Sensable Inc}, ``{OpenHaptics Toolkit 3.1 - Programmer's Guide}'',  2012.

\bibitem{bolopion2011}
A.~Bolopion,\ B.~Cagneau,\ S.~Redon,\ S.~R\'{e}gnier,  Variable gain haptic
  coupling for molecular simulation.   In  \textit{World Haptics Conference
  (WHC), 2011 IEEE}; 2011.

\bibitem{gamma1995comppatt}
E.~Gamma,\ R.~Helm,\ R.~Johnson,\ J.~Vlissides, \textit{Design Patterns -
  Elements of Reusable Object-Oriented Software;} Addison-Wesley: 2009.

\bibitem{kneller1991}
G.~R. Kneller, \textit{Mol. Simul.} \textbf{1991,} \textsl{7,} 113-119.

\end{thebibliography}

\providecommand{\refin}[1]{\\ \textbf{Referenced in:} #1}


%


\end{document}